\documentclass[12pt]{article}

\usepackage[margin=1in]{geometry}
\usepackage{graphicx}
\usepackage[textwidth=8em,textsize=small]{todonotes}

\usepackage{tikz}
\usetikzlibrary{arrows}
\usepackage{amssymb}
\usepackage{amsmath}
\usepackage{bm}
\usepackage{bbm}
\usepackage{array,multirow}
\usepackage{hyperref}
\usepackage{enumerate}
\usepackage{enumitem}
\usepackage{blkarray}
\usepackage[round]{natbib}
\usepackage{longtable}
\usepackage{soul} 
\usepackage{authblk} 

\usepackage{setspace}
\newcommand{\beq}{\begin{equation}}
\newcommand{\eeq}{\end{equation}}
\newcommand{\bei}{\begin{itemize}}
\newcommand{\eei}{\end{itemize}}
\newcommand{\ben}{\begin{enumerate}}
\newcommand{\een}{\end{enumerate}}
\newcommand{\eps}{\epsilon}

\newcommand{\indicator}[1]{{\mathbbm 1}\left(#1\right)}
\newcommand\ind{\protect\mathpalette{\protect\independenT}{\perp}}
\def\independenT#1#2{\mathrel{\rlap{$#1#2$}\mkern2mu{#1#2}}}
\newcommand{\blue}{\color{black}}

\DeclareMathOperator*{\E}{E}
\DeclareMathOperator*{\EC}{E_L}
\DeclareMathOperator*{\V}{V}
\DeclareMathOperator*{\cov}{Cov}

\DeclareMathOperator*{\expit}{expit}
\DeclareMathOperator*{\logit}{logit}

\title{Heterogeneous Indirect Effects for Multiple Mediators using Interventional Effect Models}

\author[1]{Wen Wei Loh}
\author[1]{Beatrijs Moerkerke}
\author[1]{Tom Loeys}
\affil[1]{Department of Data Analysis, Ghent University, Gent, Belgium.}
\author[2,3]{Stijn Vansteelandt}
\affil[2]{Department of Applied Mathematics, Computer Science and Statistics, Ghent University, Ghent, Belgium}
\affil[3]{Department of Medical Statistics, London School of Hygiene and Tropical Medicine, London, United Kingdom}

\date{}

\begin{document}

\maketitle

\begin{abstract}
Decomposing an exposure effect on an outcome into separate natural indirect effects through multiple mediators requires strict assumptions, such as correctly postulating the causal {\blue structure} of the mediators, and no unmeasured confounding among the mediators.
In contrast, interventional indirect effects for multiple mediators can be identified even when - as often - the mediators either have an unknown causal structure, or share unmeasured common causes, or both.
Existing estimation methods for interventional indirect effects require calculating each distinct indirect effect in turn. This can quickly become unwieldy or unfeasible, especially when investigating indirect effect measures that may be modified by observed baseline characteristics.
In this article, we introduce simplified estimation procedures for such heterogeneous interventional indirect effects using interventional effect models. Interventional effect models are a class of marginal structural models that encode the interventional indirect effects as causal model parameters, thus readily permitting effect modification by baseline covariates using (statistical) interaction terms.
The mediators and outcome can be continuous or noncontinuous.
We propose two estimation procedures: one using inverse weighting by the counterfactual mediator density or mass functions, and another using Monte Carlo integration. The former has the advantage of not requiring an outcome model, but is susceptible to finite sample biases due to highly variable weights. The latter has the advantage of consistent estimation under a correctly specified (parametric) outcome model, but is susceptible to biases due to extrapolation.
The estimators are illustrated using publicly available data assessing whether the indirect effects of self-efficacy on fatigue via self-reported post-traumatic stress disorder symptoms vary across different levels of negative coping among health care workers during the COVID-19 outbreak.
\end{abstract}



\section{Introduction}

Mediation analysis is widely used to assess the effect of an exposure or treatment ($A$) on an outcome ($Y$) that is transmitted via an intermediate variable that lies on a causal pathway from $A$ to $Y$. A formal framework for mediation analysis has been developed using counterfactual-based distribution-free definitions of {\em natural direct and indirect effects} \citep{robins1992identifiability,pearl2001direct}. These developments elucidate the ignorability assumptions needed to identify the natural (in)direct effects, and explicates decomposing the total effect into a direct and an indirect effect regardless of the statistical model. Nonlinear models for the mediator and the outcome may therefore be used in practice under this framework. 
Notwithstanding these advances, natural (in)direct effects may be uninformative of real-life interventions as it may be unfeasible to set multiple variables simultaneously to individual-specific counterfactual values \citep{didelez2006direct,petersen2006estimation,naimi2014mediation}, and it is impossible to perform experiments in which the identification assumptions for natural (in)direct effects are guaranteed to be satisfied \citep{robins2010alternative}.

The problem lies in that natural (in)direct effects are defined in terms of so-called cross-world counterfactuals \citep{robins2010alternative} involving composite, or nested, counterfactuals for the mediator and outcome. 
Identification thus demands either specifying Non-Parametric Structural Equations Model \citep{pearl2009} structures for all observed variables, or assuming independence of mediator and outcome counterfactuals under different exposures; see \citet{andrews2020insights} for a discussion of the latter assumption.
In particular, this ``cross-world'' independence assumption is violated when there are multiple mediators with one mediator affecting another mediator so that the former is an exposure-induced (also termed time-varying) confounder of the mediator-outcome relation for the latter \citep{avin2005identifiability}. 
Mediation analyses in most substantive applications involve multiple mediators, either because scientific interest is in investigating the effects transmitted via multiple candidate or putative mediators, or because certain confounders affected by exposure are concurrently perceived as competing mediators.
Extensions of mediation analysis using natural effects to settings that involve multiple mediators are therefore restricted to (fine-grained) decompositions that assume (i) the mediators to either be independent \citep{lange2013assessing,taguri2018} or conform to a correctly postulated causal structure {\blue whereby certain path-specific effects are identifiable under further assumptions} \citep{vansteelandt2012natural, daniel2015causal, steen2017flexible,albert2019gmediation}, and (ii) the mediators share no hidden (or unmeasured) common causes.
But in most realistic scenarios, the causal {\blue structure} among the mediators and the hidden common causes of the mediators are unknown, thus violating the assumptions needed for identification.

In contrast, {\em interventional (in)direct effects}, first introduced by \citet{didelez2006direct} and \citet{vanderweele2014effect} for a single mediator, are not defined in terms of cross-world counterfactuals. Unlike natural effects that are defined in terms of individual-level (deterministic) interventions on the mediator, interventional effects consider population-level (stochastic) interventions that set the value of the mediator to a random draw from its counterfactual distribution. {\blue This distinction allows identification of interventional indirect effects in the context of multiple mediators under empirically verifiable assumptions that could be ensured in a randomized trial \citep{moreno2018understanding}.}
Furthermore, the interpretation of interventional effects remains meaningful even when the exposure cannot be manipulated at the individual level, {\blue and the natural and interventional effects may coincide empirically \citep{micali2018maternal}.}
For example, \citet{jackson2018decomposition} describe interventional (in)direct effects using race as the exposure and socioeconomic status as the mediator, without having to define nested potential outcomes with respect to race or the exposure effects of race. 
\citet{vansteelandt2017interventional} generalized the definition of interventional effects to the multiple mediator setting, and demonstrated that the joint indirect effect of an exposure on an outcome can be decomposed into separate indirect effects via each distinct mediator, and an indirect effect via the mediators' mutual dependence, regardless of the underlying causal structure of the mediators. 
In particular, an interventional indirect effect via a mediator is defined to capture the combined effect along all underlying causal pathways leading from exposure (possibly via other mediators) to the mediator of interest, then from the mediator directly to the outcome. 
Interventional (in)direct effects for multiple mediators are well-defined and can be identified, and thus unbiasedly estimated, even when the directions of the causal effects between the mediators are unknown, or the mediators are manifestations of an underlying latent process, or the mediators share hidden common causes. 
Recent discussions of interventional (in)direct effects include {\blue comparisons with natural (in)direct effects} \citep{moreno2018understanding,lok2019organic,trang2019clarifying}, longitudinal mediation \citep{vanderweele2017mediation}, and path-specific interpretations for multiple mediators \citep{lin2017interventional}.

In this article, we consider heterogeneous interventional (in)direct effects for multiple mediators that can vary across different levels of the observed baseline characteristics.
We build on the framework of \citet{vansteelandt2017interventional} to develop interventional indirect effect measures that may be modified by the baseline covariates.
The mediators and outcome can be continuous or noncontinuous. 
Existing estimation methods demand either deriving closed form expressions of the estimators when the assumed mean models for the mediators and the outcome are linear, or separately calculating each (in)direct effect in turn when the assumed models are nonlinear.
When estimating interventional indirect effects that may take on different values for different levels of the observed covariate(s), either the derivations become complicated and prone to miscalculations, or the calculations have to be repeated for each unique covariate level in turn.
Existing methods for inferring heterogeneous interventional (in)direct effects can potentially be unwieldy or unfeasible in such settings.
To address this shortcoming, we propose using {\em interventional effect models} for straightforward and simultaneous modeling of the interventional (in)direct effects when there are multiple mediators. 
Interventional effect models adopt the functional form of {\em natural effect models} for multiple mediators \citep{lange2013assessing, steen2017flexible}, a class of marginal structural mean models \citep{robins2000marginal}, that express the mean potential outcomes in terms of hypothetical exposure levels for each mediator. 
Indirect and direct effects can be readily encoded as causal parameters that index these models, while allowing for effects to vary between different covariate levels via interactions between the exposure levels and the covariates.
Directly modeling the interventional (in)direct effects of interest can therefore simplify estimation and inference.
We propose two estimation procedures for fitting interventional effect models for multiple mediators: inverse weighting by the counterfactual mediator density or mass functions, or Monte Carlo integration. The key advantage of the former is that no outcome model needs to be (correctly) specified. However, this estimator can be susceptible to non-negligible finite sample biases due to highly variable weights. In contrast, the latter estimator has the advantage of consistent estimation under a correctly assumed (parametric) model for the outcome. But specifying an outcome model that is coherent with the posited effect model can be difficult or impossible for nonlinear effect models due to noncollapsibility  \citep{greenland1999confounding}. This renders the latter estimator susceptible to biases due to extrapolation.

The remainder of this article is as follows. In Section~\ref{sect:notation_definition} notation is introduced, and the heterogeneous interventional (in)direct effects are defined following \citet{vansteelandt2017interventional}. The exact decomposition of the total effect into the direct effect and indirect effects via each mediator is presented, and the existing estimation method is briefly reviewed. In Section~\ref{sect:effectmodels} interventional effect models are introduced, and the estimation procedures, using either inverse weighting (by the counterfactual mediator density or mass functions), or Monte Carlo integration, are described. 
In Section~\ref{sect:simstudies} the proposed methods are assessed via extensive simulation studies. 
In Section~\ref{sect:application} the methods are utilized to assess the effect of self-efficacy on fatigue that is possibly mediated by different post-traumatic stress disorder symptoms, and whether the indirect effects vary across different levels of negative coping, among health care workers in China during the COVID-19 outbreak \citep{hou2020selfefficacy}. 
{\blue The \texttt{R} scripts used to implement the proposed estimation procedures in carrying out the simulation studies in Section~\ref{sect:simstudies}, and the illustration in Section~\ref{sect:application}, are freely available online\footnote{\url{https://github.com/wwloh/interventional-effects}}. More user-friendly functions in an \texttt{R} package for applied researchers are under development.} A brief discussion is provided in Section~\ref{sect:discussion}.

\section{Interventional effects \label{sect:notation_definition}}

\subsection{Notation for counterfactual mediators and potential outcomes}

Consider the setting with an exposure $A$, multiple mediators $M_1, \ldots, M_t$, and an outcome $Y$. 
For $s=1,\ldots,t$, let $M_{s,a}$ denote the (individual) counterfactual for $M_s$ when $A$ is set to $a$. 
Let $\tilde M_{s,a|L}$ denote a random draw from the {\em marginal} counterfactual mediator distribution that does not depend on any other mediators (given baseline covariates $L$) when exposure $A$ is set to $a$; i.e., $\tilde M_{s,a|L} \sim F(M_{s,a}|L)$, where $F(\cdot)$ denotes a cumulative distribution function.
It is assumed that all such covariates $L$ are themselves unaffected by exposure or any mediators; otherwise they should be included among the set of mediators.
For notational simplicity, write $\tilde M_{s,a|L}$ as $\tilde M_{s,a}$ where the dependence on $L$ is implied.
{\blue Hence the counterfactual mediators $\tilde M_{s,a}=M_{s,a}$ are equivalent when the covariates $L$ are sufficiently rich so that the distribution $F(M_{s,a}|L)$ is degenerate with support (i.e., non-zero probability) at only the (counterfactual) value of $M_{s,a}$; otherwise $\tilde M_{s,a}$ will differ from $M_{s,a}$.}
Let $Y_{a m_1\cdots m_t}$ denote the potential outcome for $Y$ if, possibly counter to fact, $A$ is set to $a$, and each mediator $M_s$ is set to the value $m_s, s=1,\ldots,t$. 
For example, the potential outcome for $Y$ if exposure $A=a^{(0)}$ and when $M_s=\tilde M_{s,a^{(s)}}, s=1,\ldots,t$, is thus $Y_{a^{(0)}\tilde M_{1,a^{(1)}}\cdots\tilde M_{t,a^{(t)}}}$. 
Let $\tilde {\bm M}_{a} = (\tilde M_{1,a}, \ldots, \tilde M_{t,a})$ denote a random draw of the counterfactual mediators from their {\em joint} distribution under exposure $A=a$ (given $L$); i.e., $\tilde {\bm M}_{a} \sim F(M_{1,a},\ldots,M_{t,a}|L)$.
Here and throughout (counterfactual) mediators are written in bold if and only if referring to (values from) their joint distribution.
Note that both marginal and joint (counterfactual) mediator distributions are {\em conditional} on the baseline covariates $L$, although we omit the dependence for notational convenience.
Let $Y_{a^{(0)}\tilde {\bm M}_{a^{(1)}}}$ denote the potential outcome if exposure $A$ is set to $a^{(0)}$ and when the values of the mediators are jointly drawn from $F(M_{1,a^{(1)}},\ldots,M_{t,a^{(1)}}|L)$.

\subsection{Definition and decomposition of interventional effects}
In this section, we define the interventional (in)direct effects and describe possible decompositions for a binary exposure $A$ following \citet{vansteelandt2017interventional}.
The average potential outcomes (hereafter termed ``estimands'') among individuals with a fixed value of the baseline covariates $L=l$ are defined by:
\[
\E(Y_{a^{(0)}\tilde {\bm M}_{a^{(1)}}}|L=l)
= \int \E(Y_{a^{(0)} m_1\cdots m_t}|L=l) \; 
dF(M_{1,a^{(1)}}\!=\!m_1,\ldots,M_{t,a^{(1)}}\!=\!m_t| L=l),
\]
and
\[
\E(Y_{a^{(0)}\tilde M_{1,a^{(1)}}\cdots\tilde M_{t,a^{(t)}}}|L=l)
=\int \E(Y_{a^{(0)} m_1\cdots m_t}|L=l) \; 
dF(M_{1,a^{(1)}}\!=\!m_1| L=l) \cdots dF(M_{t,a^{(t)}}\!=\!m_t| L=l).
\]
For notational simplicity, we henceforth write the conditional expectation as $\EC(\cdot) = \E(\cdot|L)$.
Let $g(\cdot)$ denote a user-specified canonical link function. 
{\blue Let $Y_{a}$ denote the potential outcome for $Y$ if, possibly counter to fact, $A$ is set to $a$. 
Define the total effect as $g\left\{\EC\left(Y_{1}\right)\right\} - g\left\{\EC\left(Y_{0}\right)\right\}$,} which can be decomposed into a direct effect and an indirect effect, respectively:
\begin{align}
{\rm DE_L}&= 
g\left\{\EC(Y_{1\tilde {\bm M}_{1}})\right\} - g\left\{\EC(Y_{0\tilde {\bm M}_{1}})\right\}, \label{eq:DE_define}\\
{\rm IE_L}&= 
g\left\{\EC(Y_{0\tilde {\bm M}_{1}})\right\} - g\left\{\EC(Y_{0\tilde {\bm M}_{0}})\right\}. \label{eq:IE_define}
\end{align}
The direct effect describes the difference between the (transformed) estimands under different exposures, while holding the (joint) distribution of the (counterfactual) mediators fixed under exposure $A=1$, in the subgroup defined by the covariate(s) $L$.
The indirect effect is the difference between the (transformed) estimands when the (counterfactual) mediators' (joint) distribution is shifted from exposure to control, among individuals whose exposure levels are set to control in the subgroup defined by $L$.
The (joint) indirect effect \eqref{eq:IE_define} can be further decomposed into separate indirect effects via each mediator, and an indirect effect via the mutual dependence of the mediators.
Define the indirect effect via the $s$-th mediator $M_s,s=1,\ldots,t$, as: 
\beq\label{eq:IEs_define}
{\rm IE_L}_s = 
g\left\{\EC(Y_{0\tilde M_{1,1}\cdots\tilde M_{s-1,1}\tilde M_{s,1}\tilde M_{s+1,0}\cdots\tilde M_{t,0}})\right\} - 
g\left\{\EC(Y_{0\tilde M_{1,1}\cdots\tilde M_{s-1,1}\tilde M_{s,0}\tilde M_{s+1,0}\cdots\tilde M_{t,0}})\right\}.
\eeq
The exposure $a^{(s)}$ for the $s$-th mediator takes the value $1$ in the first term, and the value $0$ in the second term.
This indirect effect is therefore the difference between the (transformed) estimands when the $s$-th mediator's (marginal) distribution is shifted from one exposure level to another, while holding the distributions of all other mediators fixed among individuals with covariate value $L$ whose exposure levels are set to control.
By shifting the marginal distribution of the counterfactual mediator in question - averaged over all other mediators - this indirect effect is readily interpreted as the combined effect of exposure on the mediator, along all underlying causal pathways (possibly via other mediators), and the effect of the mediator on the outcome. 
{\blue \citet{moreno2019defining} describe how the shifts in the mediators' distributions can be interpreted as hypothetical interventions in a future ``target'' trial. Such an interpretation avoids the notion of underlying path-specific effects among the mediators, because any future (randomized) intervention on a mediator (alone) may be designed so that all other mediators are unaffected by that intervention.}
Because the (possibly arbitrary) mediator indices are used merely as labels and need not imply any assumed causal ordering, a sensitivity analysis can be readily carried out by permuting the mediator indices and assessing the indirect effects under each permutation. We conduct such a sensitivity analysis in the applied example.

The total, direct and joint indirect effects are defined using potential outcomes where the mediators are drawn from their joint (counterfactual) distribution, whereas the separate indirect effects via each mediator are defined using potential outcomes where the mediators are drawn from their respective marginal (counterfactual) distribution. 
This distinction leads to the following difference between the joint indirect and the sum of the separate indirect effects, where the latter equals:
\beq\label{eq:IE_marginal_tM}
\sum_{s=1}^{t} {\rm IE_L}_s=
g\left\{\EC(Y_{0\tilde M_{1,1}\cdots\tilde M_{t,1}})\right\} - 
g\left\{\EC(Y_{0\tilde M_{1,0}\cdots\tilde M_{t,0}})\right\}.
\eeq
In general, the sum \eqref{eq:IE_marginal_tM} need not equal the (joint) indirect effect  \eqref{eq:IE_define}. The difference is termed the indirect effect via the {\em mutual dependence of the mediators}, and is defined as:
\begin{align}
\left[g\left\{\EC(Y_{0\tilde {\bm M}_{1}})\right\} - g\left\{\EC(Y_{0\tilde {\bm M}_{0}})\right\}\right]
- 
\left[g\left\{\EC(Y_{0\tilde M_{1,1}\cdots\tilde M_{t,1}})\right\} - 
g\left\{\EC(Y_{0\tilde M_{1,0}\cdots\tilde M_{t,0}})\right\}\right]. \label{eq:IEt_mutual_define} 
\end{align}
This indirect effect is an important component in the decomposition of the joint indirect effect: it describes the mediated effect of exposure on outcome when the relationships between the mediators, and their subsequent effects on the outcome, differ for different exposure levels, so that the indirect effect via the mediators cannot be considered separately through any lone mediator. 
For example, under linear mean models for the mediators and the outcome, this indirect effect is non-zero if (i) there is non-zero mediator-mediator interaction in the outcome model; and (ii) the covariance of the mediators differs with exposure. Closed form expressions of the interventional direct and indirect effects under (correctly) assumed linear models for the mediators and the outcome when there are two mediators are provided in Appendix~\ref{sect:linear2M_POCeffects}.

\subsection{Identification of interventional effects \label{sect:identification}}

In general, the observed (baseline) covariates $L$ used to define subgroups for the heterogeneous interventional effects may be confounders of the exposure-outcome, mediator(s)-outcome, and exposure-mediator(s) relations. Again, we note that all confounders in $L$ are assumed to be unaffected by the exposure or any mediators, otherwise any exposure-induced confounders should be listed as additional mediators.
Identification of the interventional effects defined above therefore requires the following assumptions \citep{vansteelandt2017interventional}:
\bei
\item the effect of exposure $A$ on outcome $Y$ is unconfounded conditional on $L$, i.e., 
\beq
Y_{a m_1\cdots m_t} \ind A | L \quad \forall \; a, m_1, \ldots, m_t; \label{eq:identify_1}
\eeq
\item the effect of all mediators $M_1, \ldots M_t$ on outcome $Y$ is unconfounded conditional on $A$ and $L$, i.e.,
\beq
Y_{a m_1\cdots m_t} \ind (M_1,\ldots,M_t) | (A=a, L)  \quad \forall \; a, m_1, \ldots, m_t; \label{eq:identify_2}\\
\eeq
\item the effect of exposure $A$ on all mediators is unconfounded, i.e.,
\beq
(M_{1,a},\ldots,M_{t,a}) \ind A | L \quad \forall \; a. \label{eq:identify_3}
\eeq
\eei
{\blue Two additional assumptions ensure that the counterfactual mediators and potential outcomes can be identified from observed data.
The first is that {\em causal consistency} holds, i.e., that the counterfactual mediators and potential outcomes equal their observed values when the hypothetical exposure and mediator values correspond to their observed values; i.e., $M_s = M_{s,A}, s= 1,\ldots,t$, and $Y=Y_{A M_1\cdots M_t}$.
The second is that {\em positivity} holds, i.e., that each individual has (i) non-zero probabilities of being assigned to either the exposure or control group, given the covariates, e.g., $\Pr(A=a|L)>0, a=0,1$; and (ii) non-zero mediator densities, conditional on treatment and covariates, e.g., 
$f(M_{1}\!=\!m_1,\ldots,M_{t}\!=\!m_t| A=a, L) > 0$ for all values of $a, m_1, \ldots, m_t$.
}

\subsection{Existing estimation methods}

In this section, we briefly review existing estimation methods for interventional (in)direct effects as described in \citet{vansteelandt2017interventional}.
Estimation requires specifying an appropriate mean model for the outcome and a model for the mediators' joint distribution. 
The outcome mean model is conditional on the mediators $M_1,\ldots,M_t$, exposure $A$ and all covariates $L$. The model for the mediators' joint distribution is conditional on the exposure $A$ and all covariates $L$; the (implied) model for each mediator's marginal distribution can be obtained by (numerically) averaging over the distribution of the other mediators.
For each specific interventional (in)direct effect in turn, the (counterfactual) mediators are randomly sampled from their distributions under the corresponding exposure levels. The potential outcomes are predicted given the sampled mediator values using the fitted model. Monte Carlo integration is carried out by making repeated stochastic draws of the mediators, then averaging over the predicted potential outcomes across all mediator draws.
However, to estimate the aforementioned heterogeneous indirect effects, the potential outcomes would be averaged only among the subset of individuals who share the same value of the observed covariate(s) $L$.
Repeating this calculation for each subgroup in turn can quickly become unwieldy in practice, especially when there are more than two mediators or more than a few unique covariate values, or both. 
Furthermore, it is unfeasible to carry out these calculations when the covariates are continuous and only a few individuals are in a subgroup defined by the same covariate value(s).
In the next section, we propose an approach to simplify estimation of heterogeneous interventional effects.

\section{Interventional effect models \label{sect:effectmodels}}

\subsection{Effect models}
We now describe the interventional effect models that facilitate estimating the interventional (in)direct effects proposed in this article. 
We adopt the same functional form as natural effect models \citep{lange2012simple, steen2017flexible} that generalize marginal structural models \citep{robins2000marginal} to express the mean potential outcomes in terms of hypothetical exposure levels for each mediator. 
The parameters in natural effect models describe differences between the estimands that exactly encode the natural (in)direct effects.
But unlike natural effect models that focus on decomposing natural indirect effects, we will use interventional effect models to parametrize the aforementioned interventional (in)direct effects so as to allow simultaneous estimation.

For pedagogical purposes, we will consider the setting with two mediators $M_1$ and $M_2$, and two covariates $L_1$ and $L_2$, and defer models for settings with more mediators and covariates to the simulation studies and the applied example.
An (interventional) effect model that encodes (in)direct effects that are modified only by the covariate $L_1$ is:
\begin{align}
&g\left\{\EC\left(Y_{a m_1 m_2}\right)\right\} \notag\\
&= \mu_{0} + 
\left\{ \theta_{1}a^{(1)} + \theta_{2}a^{(2)} + \theta_{1c}a^{(1)}L_1 + \theta_{2c}a^{(2)}L_1 \right\} (1-J) + \mu_{1}L_1 + \mu_{2}L_2 \notag\\ 
&\quad+ \left\{ \mu_{0J} + \gamma_0 a^{(0)} + \gamma_{0c} a^{(0)}L_1
+ \gamma_1 a^{(1)}\indicator{a^{(1)}=a^{(2)}} + \gamma_{1c} a^{(1)}\indicator{a^{(1)}=a^{(2)}}L_1 \right\}J.
\label{eq:effects_model_a00}
\end{align}
{\blue For notational simplicity here and throughout, we will use the subscripts on the left hand side of \eqref{eq:effects_model_a00} merely to emphasize that the potential outcomes for $Y$, and not the observed outcomes $Y$, are the dependent variables in the effect models.}
The indicator function $\indicator{B}$ takes value 1 when event $B$ is true, or $0$ otherwise.
The interactions between the mediators' exposure levels and the covariate $L_1$ allow for (in)direct effects to vary across different values of $L_1$. (The subscripts $c$ in the coefficients $\theta_{1c}, \theta_{2c}, \gamma_{0c},\gamma_{1c}$ emphasize the interactions that encode the effect modification by the covariate of interest.)
The indicator $J$ takes value $0$ when the potential outcomes are defined using mediators that are randomly drawn from their respective marginal (counterfactual) distributions, or $1$ when drawn from their joint (counterfactual) distribution.
When $J=0$, the indirect effect via $M_1$ is encoded by $(\theta_{1}+ \theta_{1c}L_1)$ for some fixed value of $L_1$.
Similarly, the indirect effect via $M_2$ is encoded by $(\theta_{2}+ \theta_{2c}L_1)$ for some fixed value of $L_1$.
The indirect effect measures are therefore modified by the covariates when the parameters $\theta_{1c}$ or $\theta_{2c}$ are non-zero.
The sum of the indirect effects via both mediators is readily obtained by summing the parameters $\sum_{s=1}^{2}(\theta_{s}+ \theta_{sc}L_1)$.
The joint indirect effect is defined using potential outcomes where the mediators are drawn from their joint (counterfactual) distribution under exposure level $A=a^{(1)}$, hence restricting the exposure levels for both mediators to be the same when $J=1$, i.e., $a^{(1)}=a^{(2)}$. The joint indirect effect is thus encoded by $(\gamma_1+\gamma_{1c}L_1)$ for some fixed value of $L_1$.
The indirect effect via the mutual dependence of the mediators is the difference between the joint indirect effect and sum of the separate indirect effects, and is thus encoded by $(\gamma_1+\gamma_{1c}L_1) - \sum_{s=1}^{2}(\theta_{s}+ \theta_{sc}L_1)$.
There is no main effect for the exposure $a^{(0)}$ alone when $J=0$ because the indirect effects defined in this paper are fixed at $a^{(0)}=0$.
The direct effect is encoded by the parameters $(\gamma_0+\gamma_{0c}L_1)$ for some fixed value of $L_1$ when $J=1$.
Lastly, the total effect is the sum of the direct and joint indirect effects, which is simply the sum of the parameters $\left(\sum_{j=0}^{1} \gamma_j + \gamma_{jc} \right)$. 
Because interventional effect models are conditional mean models for the potential outcomes, main effects for both $L_1$ and $L_2$ are included in the effect model to adjust for confounding, even when the association between $L_2$ and the (potential) outcome is not of primary interest \citep{vansteelandt2012imputation}.
Interaction terms $J \times L_1$ and $J \times L_2$ may also be included in the effect model to allow the effect of each confounder on the (potential) outcome to differ depending on whether the (counterfactual) mediators are randomly drawn from either their respective marginal distributions $(J=0$), or their joint distribution $(J=1$).

\subsection{Estimation via inverse weighting}

Estimators of the interventional direct and indirect effects can be obtained by fitting the effect model using weights that are (inversely) proportional to the counterfactual mediator density or mass functions. 
For each individual, duplicated data is constructed using different levels of the exposure $a^{(0)}, a^{(1)}, a^{(2)}$, under either marginal ($J=0$) or joint ($J=1$) distributions, for the counterfactual mediators.
The posited effect model
is then fitted to the duplicated data using weighted regression. The procedure is as follows:

\ben[label=A\arabic*.]
\item Fit a propensity score model for exposure conditional on all observed baseline confounders to the observed data. 
This is required to adjust for exposure-outcome and exposure-mediator(s) confounding toward satisfying the identifying assumptions \eqref{eq:identify_1} and \eqref{eq:identify_3}.
For example, a logistic regression model may be:
\[
\logit\{\Pr(A=1|L)\} = \beta_0 + \beta_l L.
\]
Let $\hat p = \expit(\hat\beta_0 + \hat\beta_l L)$ denote the (individual) predicted probability of receiving the observed exposure based on the maximum likelihood estimates of the parameters $\hat\beta_0$ and $\hat\beta_l$.
Calculate the estimated weight for each individual as $\hat w^a = A/\hat p + (1-A)/(1-\hat p)$.

\item Within each observed exposure group $A=a$, fit a group-specific model for the joint density or mass function of the (counterfactual) mediators, conditional on all observed baseline confounders $L$, to the observed data.  The confounders $L$ are included in the mediator models to adjust for mediator(s)-outcome confounding toward satisfying the identifying assumption \eqref{eq:identify_2}.
Denote the resulting estimated density by $\hat f^a(\bm M |L),a=0,1$. For example, suppose that $M_1$ is a binary mediator and $M_2$ is a normally-distributed mediator. The joint density can be factorized as $\hat f^a(\bm M |L) = \hat f^a(M_2 |M_1,L) \hat f^a(M_1 |L)$. 
A logistic regression model for $M_1$ among individuals with exposure $A=a$ may be:
\[
\logit\{\Pr(M_1=1|A=a,L)\} = \alpha_{10}^a + \alpha_{1l}^a L,
\]
where $\logit(x) = \log(x)/\log(1-x)$ and the superscript $a$ in the regression coefficients denote their dependence on $A=a$.
The implied estimated distribution is therefore a Bernoulli distribution with probability of success $\hat\E^a(M_1|L)=\expit(\hat\alpha_{10}^a + \hat\alpha_{1l}^a L)$, where $\hat\alpha_{10}^a$ and $\hat\alpha_{1l}^a$ are the maximum likelihood estimates (MLE) of the parameters, and $\expit(x)=\exp(x)/\{1+\exp(x)\}$. A linear regression model for $M_2$, conditional on $M_1$ and $L$, among individuals with exposure $A=a$ may be:
$$
{\rm E}^a(M_2 |M_1,L) = \alpha_{20}^a + \alpha_{21}^a M_1 + \alpha_{2l}^a L, \quad
{\rm V}^a(M_2 |M_1,L) = \left(\sigma_2^a\right)^2,
$$
where $\E^a(X|U)$ and $\V^a(X|U)$ respectively denote the (conditional) expectation and variance of a random variable $X$ given $U$ and exposure $A=a$.
The implied estimated distribution is therefore a Normal distribution with mean 
$\hat\alpha_{20}^a + \hat\alpha_{21}^a M_1 + \hat\alpha_{2l}^a L$ and (constant) variance $\left(\hat\sigma_2^a\right)^2$, where $\hat\alpha_{20}^a, \hat\alpha_{21}^a, \hat\alpha_{2l}^a$ and $\left(\hat\sigma_2^a\right)^2$ are the MLE of the parameters.

Let $\hat f^a(M_s |L),a=0,1$, denote the marginal density of each mediator $M_s, s=1,2$, unconditional on all other mediators but given the baseline confounders $L$, as implied by the mediators' joint distribution $\hat f^a(\bm M |L)$. 
Continuing the above example, it follows from the laws of total expectation and of total variance that $M_2$ given $L$ is Normally distributed with mean 
$\hat\alpha_{20}^a + \hat\alpha_{21}^a \hat\E^a(M_1|L) + \hat\alpha_{2l}^a L$ and (constant) variance $\left(\hat\sigma_2^a\right)^2 + \left(\hat\alpha_{21}^a\right)^2 \hat\V^a(M_1|L)$, where $\hat\V^a(M_1|L)=\hat\E^a(M_1|L)\{1-\hat\E^a(M_1|L)\}$ for binary $M_1$.

\item Construct the duplicated data for each individual as shown in Table~\ref{table:duplicatedall_inverseweighting}. 
In the first row, set all the hypothetical exposure levels to 0; e.g., $a^{(0)}=a^{(1)}=\cdots=a^{(t)}=0$.
In each row $s=2,\ldots,t+1$, set the hypothetical exposure levels as $a^{(k)}=1$ if $k\in\{1,\ldots,s-1\}$, or $a^{(k)}=0$ otherwise.
The difference between the estimands in rows $s$ and $s-1$ therefore corresponds to the interventional indirect effect via the mediator $M_{s-1}$; e.g., the indirect effect via $M_2$ corresponds to the difference between the estimands in the third and second rows. 
Similarly, the difference between the estimands in rows $1$ and $t+1$ corresponds to the sum of the interventional indirect effects via the distinct mediators. 
In the last two rows, set the hypothetical exposure levels $a^{(0)}$ to the observed value $A$; in the penultimate row set $a^{(1)}=\ldots=a^{(t)}=1-A$, and in the last row set $a^{(1)}=\ldots=a^{(t)}=A$. 
There are therefore a total of $t+3$ rows for $t$ mediators.

\begin{table}
\caption{Duplicated data for each individual with five rows when there are two mediators for a binary exposure $A$, used to estimate the parameters in an (interventional) effect model. The (counterfactual) mediators are randomly drawn from either their respective marginal distributions ($J=0$) or their joint distribution ($J=1$). The observed covariates $L$ (including $L$) are omitted for simplicity.\label{table:duplicatedall_inverseweighting}}
\centering
\renewcommand{\arraystretch}{1.5}
\begin{tabular}{ |ccc|cccc|c|} 
 \hline
 $a^{(0)}$ & $a^{(1)}$ & $a^{(2)}$
 & $J$
 & \multicolumn{3}{c|}{$\hat w_i^m(a^{(1)},\ldots,a^{(t)},J)$} &
 $Y$ \\
 \hline 
 $0$ & $0$ & $0$ & 0 & $\hat f^{0}(M_1 |L)\hat f^{0}(M_2 |L)$ & $/$ & $\hat f^A(\bm M |L)$ 
 & $Y$ \\  
 $0$ & $1$ & $0$ & 0 & $\hat f^{1}(M_1 |L)\hat f^{0}(M_2 |L)$ & $/$ & $\hat f^A(\bm M |L)$
 & $Y$ \\  
 $0$ & $1$ & $1$ & 0 & $\hat f^{1}(M_1 |L)\hat f^{1}(M_2 |L)$ & $/$ & $\hat f^A(\bm M |L)$
 & $Y$ \\  
 $A$ & $1\!-\!A$ & $1\!-\!A$ & 1 & $\hat f^{1-A}(\bm M |L)$ & $/$ & $\hat f^A(\bm M |L)$
 & $Y$ \\  
 $A$ & $A$ & $A$ & 1 & \multicolumn{3}{c|}{$1$}
 & $Y$ \\  
 \hline
\end{tabular}
\end{table}

\item 
For each row in the duplicated data, calculate the weight:
\begin{align*}
\hat w_i(a^{(0)},\ldots,a^{(t)},J) &= \indicator{a^{(0)}=A}\hat w_i^a \hat w_i^m(a^{(1)},\ldots,a^{(t)},J), \\
\mbox{where} \quad
\hat w_i^m(a^{(1)},\ldots,a^{(t)},J) &=
\left\{
\begin{array}{ll}
\prod_{s=1}^{t} \hat f^{a^{(s)}}(M_s |L) / \hat f^A(\bm M |L), & J=0; \\[12pt]
\hat f^{a^{(1)}}(\bm M |L) / \hat f^A(\bm M |L), & J=1.
\end{array}
\right.
\end{align*}

\item Fit the effect model to the duplicated data in Table~\ref{table:duplicatedall_inverseweighting} with the (observed) outcomes and computed weights from the previous step using weighted regression. 

\een
Nonparametric boostrap confidence intervals \citep{efron1994introduction}  may be constructed by randomly resampling observations with replacement and repeating all the steps for each bootstrap sample.

In step A2, separate exposure group-specific mediator models may potentially be simpler to specify (correctly) than models with interaction term(s) between the covariate(s) and exposure. 
Furthermore, it allows for exposure effects on the (joint) mediators' distributions (including their residual variances), and not merely their means (as implied by interaction terms in a single mediator model for both exposure groups).
We emphasize that the factorization of the mediators' joint density is not predicated on any (correctly) assumed causal ordering among the mediators; e.g., $\hat f^a(\bm M |L) = \hat f^a(M_1 |M_2,L) \hat f^a(M_2 |L)$ may be merely used instead.
The proposed estimation procedure follows \citet{lange2013assessing}, by fitting the effect model to the observed outcomes using weighted regression techniques. 
{\blue Consistent estimation is similarly predicated on correctly specifying (i) a propensity score model that converges in probability to its true value, and (ii) models for the (joint) distribution of the mediators, conditional on exposure and covariates, that are unbiased for the counterfactual distribution, i.e., $f^a(\bm M |L)$ converges in probability to its counterfactual distribution $f(\bm M_a |L)$, for $a=0,1$. A proof, which includes the motivation for the form of the weights, is provided in Appendix~\ref{sect:consistency}.}
However, inverse weighting by the joint mediator density or mass functions \citep{hong2010ratio,lange2012simple,steen2017flexible} may result in highly variable or volatile weights that induce estimators with non-negligible finite sample bias under certain situations. In particular, when the mediators are (strongly) associated with one another, given $A$ and $L$, due to either causal effects on each other or unobserved confounding, or both, estimators of the separate indirect effects through the mediators may be biased. We demonstrate the biases empirically in a simulation study.

\subsection{Estimation via Monte Carlo integration \label{sect:MCprocedure}}

To avoid potential biases resulting from less stable or volatile weights, we propose a second estimator that does not require inverse weighting. 
Instead, Monte Carlo draws of the counterfactual mediators are made at different exposure levels in a duplicated dataset similar to Table~\ref{table:duplicatedall_inverseweighting}.
A specified outcome model is used to predict potential outcomes given the counterfactual mediators and observed confounders in the duplicated data.
Estimators of the interventional (in)direct effects are then obtained by fitting the posited effect model to the duplicated data, using either ordinary least squares or maximum likelihood estimation. The procedure is as follows:

\ben[label=B\arabic*.]
\item Within each observed exposure group $A=a$, fit a group-specific outcome model conditional on all mediators and observed baseline confounders, e.g., $\E(Y|A=a,M_1,M_2,L)$, to the observed data. Each separate outcome model can be expressed as a function of its inputs, e.g., $\E(Y|A=a,M_1=m_1,M_2=m_2,L)=h^a(m_1,m_2,L)$, where $h^a(\cdot)$ is a user-specified function with the superscript $a$ denoting its dependence on $A=a$. Denote the estimated functions by $\hat h^a(m_1,m_2,L), a=0,1$.
For example, a logistic regression model for a binary outcome among individuals with exposure $A=a$ is:
\[
\logit\{\Pr(Y=1|A=a,M_1,M_2,L)\} = \beta_0^a + \beta_1^a M_1 + \beta_2^a M_2 + \beta_{1l}^a M_1L + \beta_{2l}^a M_2L + 
\beta_l^a L.
\]

\item Following step A2, within each observed exposure group $A=a$, fit a group-specific model for the marginal density of each mediator $M_s, s=1,2$, conditional on all baseline confounders $L$, to the observed data. Denote the resulting estimated distribution by $\hat F^a(M_s|L)$. 

\item Construct the duplicated data for each individual as shown in Table~\ref{table:duplicatedall}, which differs from Table~\ref{table:duplicatedall_inverseweighting} only in the last two rows. 
In the last two rows of Table~\ref{table:duplicatedall}, set the hypothetical exposure levels $a^{(1)}=\ldots=a^{(t)}$ to the observed value $A$; in the penultimate row set $a^{(0)}=1-A$, and in the last row set $a^{(0)}=A$. There are therefore a total of $t+3$ rows for $t$ mediators.

\begin{table}
\caption{Duplicated data for each individual with five rows when there are two mediators for a binary exposure $A$, used to estimate the parameters in an (interventional) effect model. The (counterfactual) mediators are randomly drawn from either their respective marginal distributions ($J=0$) or their joint distribution ($J=1$). The asterisk denotes a stochastically imputed counterfactual mediator. The observed covariates $L$ (including $L$) are omitted for simplicity.\label{table:duplicatedall}}
\centering
\renewcommand{\arraystretch}{1.5}
\begin{tabular}{ |ccc|ccc|c|} 
 \hline
 $a^{(0)}$ & $a^{(1)}$ & $a^{(2)}$
 & $J$
 & $\tilde M_{1,a^{(1)}}$ & $\tilde M_{2,a^{(2)}}$ &
 $h^{a^{(0)}}(m_1\!=\!M_{1,a^{(1)}},m_2\!=\!M_{2,a^{(2)}},L)$ \\
 \hline 
 $0$ & $0$ & $0$ & 0 & $M_{1,0}^\ast$ & $M_{2,0}^\ast$ 
 & $\hat h^{0}(m_1\!=\!M_{1,0}^\ast,m_2\!=\!M_{2,0}^\ast,L)$ \\ 
 $0$ & $1$ & $0$ & 0 & $M_{1,1}^\ast$ & $M_{2,0}^\ast$
 & $\hat h^{0}(m_1\!=\!M_{1,1}^\ast,m_2\!=\!M_{2,0}^\ast,L)$ \\ 
 $0$ & $1$ & $1$ & 0 & $M_{1,1}^\ast$ & $M_{2,1}^\ast$
 & $\hat h^{0}(m_1\!=\!M_{1,1}^\ast,m_2\!=\!M_{2,1}^\ast,L)$ \\ 
 $1\!-\!A$ & $A$ & $A$ & 1 & \multicolumn{2}{c|}{${\bm M}=(M_1,M_2)$}
 & $\hat h^{1-A}(m_1\!=\!M_{1},m_2\!=\!M_{2},L)$ \\ 
 $A$ & $A$ & $A$ & 1 & \multicolumn{2}{c|}{${\bm M}=(M_1,M_2)$}
 & $Y$ \\  
 \hline
\end{tabular}
\end{table}

\item For each row $s=1,\ldots,t+1$ where $J=0$, randomly sample the counterfactual mediator values (denoted by asterisks in the superscripts) from their respective (estimated) marginal distributions $\hat F^{a^{(s)}}(M_s|L)$.

\item Using the separate fitted outcome models for $A=1$ and $A=0$, impute the expected potential outcomes in each row as predictions $\hat h^{a^{(0)}}(m_1=\tilde M_{1,a^{(1)}},m_2=\tilde M_{2,a^{(2)}}, L)$, depending on whether $a^{(0)}=0$ or $a^{(0)}=1$. In the last row where $a^{(0)}=A$ and $J=1$, set the potential outcome to its observed value.

\item Fit the effect model to the duplicated data for the observed sample. 
\een
Nonparametric boostrap confidence intervals may be constructed by randomly resampling observations with replacement and repeating all the steps for each bootstrap sample.

No (propensity score) model for the exposure (given the observed confounders $L$) is required, even when exposure is non-randomly assigned, because including the observed confounders $L$ in the outcome and mediator models suffices to adjust for exposure-outcome and exposure-mediator(s) confounding toward satisfying the identifying assumptions \eqref{eq:identify_1} and \eqref{eq:identify_3}.
In step B1, separate exposure group-specific outcome mean models are recommended for the same reason as separate exposure group-specific mediator models in step A2.
The models can potentially be simpler to specify (correctly) than a single model that includes covariate-exposure and (possibly numerous) exposure-mediator(s) interaction effect(s) on the observed outcome.
Assuming that the residuals from the mediator models fitted to the observed data in step B2 are independent of all variables, an alternative to random draws of the mediators from specified parametric distributions in step B4 is to resample (with replacement) the residuals, then attach them to the predicted (counterfactual) mediators. 
{\blue In the last two rows of Table~\ref{table:duplicatedall}, the observed values of the mediators are viewed as being drawn from their (possibly unknown) joint distribution under the observed exposure level. Using the observed values has the benefit of being computationally efficient (by avoiding sampling the counterfactual mediators). Furthermore, the risks of biases due to model misspecification can be avoided, because assuming a model for the joint distribution of the (counterfactual) mediators, conditional on the confounders and (hypothetical) exposure, is unnecessary for the proposed Monte Carlo estimator.}
Because our interest is only in the expected potential outcome in each row, steps B4 and B5 may be repeated e.g., 100 times, to average over the (counterfactual) distributions of the mediators using Monte Carlo integration.
In each row of the duplicated data, the averaged (predicted) potential outcomes over all Monte Carlo draws can then be employed as the imputed potential outcome for that individual.

The proposed imputation-based procedure follows the estimation strategy of \citet{vansteelandt2012imputation} for a single mediator. It is similar to the imputation-based strategies for G-computation \citep{snowden2011implementation}, and shares their virtues and limitations \citep{vansteelandt2011invited}. 
Consistent estimation is predicated on correctly specifying (i) an outcome model conditional on exposure, mediators, and covariates that is unbiased for the marginal structural mean model, i.e., 
$h^a(m_1, \ldots, m_t,L)$ converges in probability to $\E\left(Y_{a m_1\cdots m_t}|L\right)$;
and (ii) models for the (marginal) distributions of the mediators, conditional on exposure and covariates, that are unbiased for their counterfactual distributions, i.e., for $s=1,\ldots,t$, $f^a(M_{s} |L)$ converges in probability to its counterfactual distributions $f(M_{s,a} |L)$, for $a=0,1$. A proof is provided in Appendix~\ref{sect:consistency}.
However, its simplicity may belie the same difficulty facing multiple imputation estimators for missing data analyses: specifying a model for the outcome that is ``congenial'' \citep{meng1994multiple}, or ``coherent,'' with the effect model of interest. An assumed (parametric) outcome model that is uncongenial with the posited effect model may result in extrapolation (or ``misspecification'') bias, although uncongeniality can be avoided in principle by using saturated or rich outcome models \citep{vansteelandt2012imputation}.
For example, when the effect model is linear, i.e., $g(\cdot)$ being the identity link, an outcome model that reflects the structure of the effect model may be obtained by (i) considering only the submodel where $J=0$, (ii) replacing $a^{(s)}$ with $M_s, s=1,\ldots,t$, then (iii) adding mediator-mediator(-covariate) interaction terms, such as $M_1 \times M_2$ and $M_1 \times M_2 \times L$, as well as an exposure $A$ and covariate-exposure interaction(s) $A \times L$.
We provide such an example in the illustration.
But in practice, it may be unfeasible to fit to the observed data saturated models when there are continuous covariates or mediators, or models with all possible interaction and higher-order terms when there are high-dimensional (continuous) covariates or mediators.
Furthermore, specifying an outcome model with the correct structure as the posited effect model can be difficult when the latter is nonlinear due to noncollapsibility; see e.g., \citet{vanderweele2010odds} and \citet{tchetgen2014note} when using logistic regression for binary outcomes. 
{\blue We evaluate the impact of misspecifying the outcome model on the effect estimates empirically in a simulation study, and defer formal assessment of the sensitivity of the estimators to model misspecification to future work.}

\section{Simulation studies \label{sect:simstudies}}

{\blue
Three simulation studies across 45 different settings were conducted to empirically assess the operating characteristics of the proposed estimators using either inverse weighting or Monte Carlo integration. Details of the data-generating procedures are deferred to Appendix~\ref{sect:simstudies_details}. 
To provide an overview, in study 1, we considered a simple setting with two mediators and a single confounder, with no interaction terms in the data-generating models, so that the (in)direct effects were the same across different levels of the confounder. The objective of this study was to compare the empirical biases of the two estimation procedures, especially the impact of the (in)stability of the weights for the inverse weighting estimator on any potential biases. 
To demonstrate that the mediators' joint density used in the inverse weighting estimator did not require correctly ordering the mediators (e.g., $M_1 \rightarrow M_2$), we simply factorized the joint density into its conditional distributions using an incorrect causal ordering instead (e.g., $M_2 \rightarrow M_1$).
In study 2, we built on study 1 to introduce an additional confounder, and a exposure-confounder interaction effect on a mediator, in the data-generating model to render separate indirect effects via each mediator that differed across different levels of that confounder. In addition to the empirical biases of the two estimation procedures, we compared the empirical coverage of their corresponding nonparametric bootstrap percentile 95\% confidence intervals.
In study 3, we built on both studies 1 and 2 by further including exposure-mediator and mediator-mediator interaction effects in the data-generating model. This resulted in all the indirect effects, including the effect via the mediators' mutual dependence, being modified by a confounder. In addition to assessing the empirical biases of both the inverse weighting and Monte Carlo estimators under correctly specified models, we considered a third Monte Carlo estimator that misspecified the outcome model. In all three studies we considered settings with two (normally-distributed) mediators, where one may have been affected by the other, and with an outcome that was binary in studies 1 and 3, and normally-distributed in study 2. 

We summarize the results of each simulation study comparing the two proposed estimation procedures, and refer readers to the detailed findings in Appendix~\ref{sect:simstudies_details}.
In study 1, the inverse weighting estimators of the direct and joint indirect effects were empirically unbiased under all the considered settings, even when an incorrect ordering for factorizing the joint mediator density was chosen. 
The estimators of the separate indirect effects via each mediator were empirically unbiased, but only when either (i) there were no or weak causal effects among the mediators, (ii) the effect of exposure transmitted (indirectly) to the mediator in question via any other mediators was similar to the effect transmitted (directly) to the mediator. 
The non-negligible biases, which persisted even at a large sample size, were due to less stable weights when one of the mediators was strongly affected (indirectly only) by exposure. 
In contrast, the Monte Carlo estimators were empirically unbiased under all the considered settings. 
Note that the Monte Carlo estimators as proposed in this paper requires no assumptions about the causal ordering of the mediators, because models for only the marginal distribution of each mediator (unconditional on all other mediators), and not the mediators' joint distribution, are required.
In study 2, similar results for the inverse weighting point estimates were arrived at. Furthermore, the instability in the weights led to (possibly severe) under-coverage of the confidence intervals (far) below their nominal levels.
In contrast, estimators using the Monte Carlo procedure were empirically unbiased, and the confidence intervals approximately attained the nominal coverage, under all considered settings.
In study 3, similar results for the inverse weighting point estimates were again arrived at. 
However, when the fitted outcome model incorrectly included only main effects for the mediators, the Monte Carlo estimators were biased even at a large sample size.
When the outcome model (correctly) included the mediator-mediator interaction terms, the Monte Carlo estimators displayed finite sample biases and larger variability empirically that were reduced as the sample size increased.

In conclusion, the inverse weighting estimators were susceptible to empirically biased point estimates and confidence intervals, due to less stable weights when one of the mediators was strongly affected (indirectly only) by exposure, even at a large sample size. In contrast, the Monte Carlo estimators were empirically consistent for their true effects when the (parametric) outcome model was correctly specified.
}

\section{Application \label{sect:application}}


\subsection{Data and variables}

The proposed estimation procedures are illustrated by reanalyzing a publicly available data set\footnote{\url{https://www.openicpsr.org/openicpsr/project/119156/version/V1/view}} from an observational study exploring the association between self-efficacy and fatigue among $n=527$ health care workers during the COVID-19 outbreak \citep{hou2020selfefficacy}.
The data was collected between March 13 and 20, 2020, from a cross-sectional survey of workers in Anqing City, Anhui Province, China, which borders Hubei province, the epicenter of the COVID-19 outbreak.
Frontline health care workers experience the recurring strains and grief of treating patients during the pandemic which severely affects their mental health \citep{npr2020pandemic}.
\citeauthor{hou2020selfefficacy} investigated whether the effect of self-efficacy on fatigue among these workers could be mediated by their self-reported post-traumatic stress disorder (PTSD) symptoms.
We describe the variables merely for illustrating the proposed interventional (in)direct effects and estimation procedures in this article. Readers interested in the study details and substantive justification of the self-reported PTSD symptoms as a (set of) mediator(s) lying on the causal path between self-efficacy and fatigue are referred to \citeauthor{hou2020selfefficacy}.
Self-efficacy was measured by 10 items each scored on a 4-point Likert scale from 1 (not true at all) to 4 (exactly true). Fatigue (outcome $Y$) was measured by 14 dichotomous items, each indicating the presence of a symptom of physical or mental fatigue (1 if having the symptom, 0 if no symptom reported). 
The PTSD symptoms were measured by 17 items, each rated on a 5-point Likert scale from 1 (not severe at all) to 5 (extremely severe). The items for the PTSD symptoms could be classified into three subscales (re-experiencing, avoidance and hyperarousal), although \citeauthor{hou2020selfefficacy} defined a single mediator using the sum scores for all 17 items. \citeauthor{hou2020selfefficacy} further assessed whether indirect effects varied across different levels of negative coping, which was measured by 8 items each rated on a 4-point Likert scale from 0 (never used) to 3 (often used).

\begin{figure}
    \centering
\begin{tikzpicture}[->,>=stealth',shorten >=1pt,auto,node distance=5cm,thick,
  main node/.style={rectangle,fill=blue!20,draw,minimum size=1cm}]

  \node[main node] (A) {Self-efficacy ($A$)};
  \node[main node] (M2) [right of=A] {Avoidance ($M_2$)};  
  \node[main node] (M1) [above=0.75cm of M2] {Re-experiencing ($M_1$)};
  \node[main node] (M3) [below=0.75cm of M2] {Hyperarousal ($M_3$)};
  \node[main node] (L) [left of=M1] {Covariates ($L$)};  
  \node[main node] (Y) [right of=M2] {Fatigued ($Y$)};
  
  \path[black]
  	(A) edge (M1)
  	(A) edge (M2)
  	(A) edge (M3)
  	(M1) edge (Y)
  	(M2) edge (Y)
  	(M3) edge (Y)
  	(A) edge[bend right = 55] (Y)
  	;	
  \path[gray, dashed]
   	(L) edge (A)
   	(L) edge (M1)
   	(L) edge (M2)		
  	(L) edge (M3)
  	(L) edge[bend left = 40] (Y)	
  	;
\end{tikzpicture}
    \caption{Posited causal diagram for the variables in the applied example. For simplicity, the baseline covariates $L$ are represented by a single node, but the effects emanating from each separate variable in $L$ were allowed to have different strengths. 
    Under the identification assumptions \eqref{eq:identify_1}--\eqref{eq:identify_3}, the covariates in $L$ were sufficient to adjust for all confounding between $A, M_s$, and $Y$ for $s=1,2,3$.
    Notwithstanding the absence of edges among the mediators $(M_1, M_2, M_3)$, the mediators may be associated due to an unknown causal ordering, or unobserved common causes, or both.
    Edges emanating from $L$ are drawn as as gray broken lines for visual clarity. \label{fig:data}}   
\end{figure}
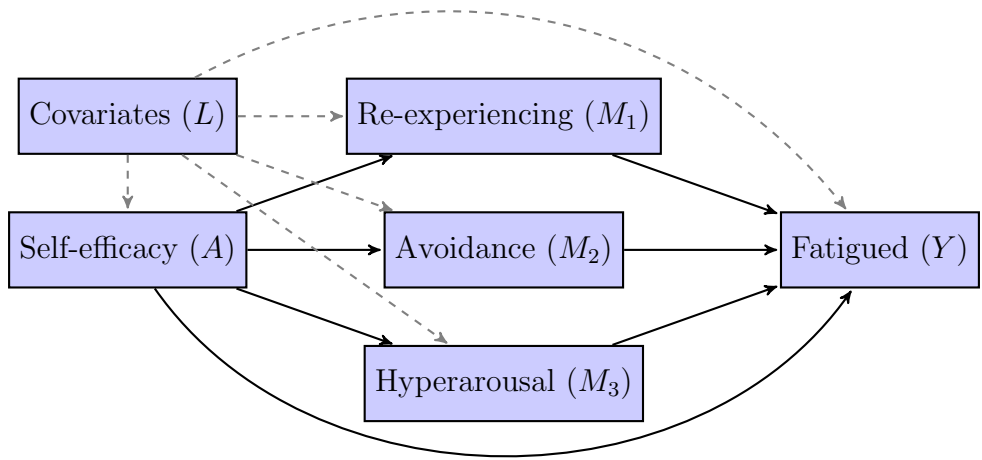

For the sole purposes of illustration, we dichotomized the sum score for self-efficacy at the empirical median by {\blue defining $A=1$ if self-efficacy was above the sample median, or 0 otherwise}. Following the recommendation by \citeauthor{hou2020selfefficacy}, we dichotomized the sum score for fatigue so that individuals were deemed to be fatigued ($Y=1$) if they reported having at least seven symptoms, or not fatigued ($Y=0$) if they reported fewer than seven symptoms.
In this paper, we considered each PTSD subscale as a separate mediator that was measured by their respective sum of the manifest items (i.e., ``sum score''). The three (continuous) mediators were therefore re-experiencing ($M_1$; 5 items), avoidance ($M_2$; 7 items) and hyperarousal ($M_3$; 5 items).
We defined the sum score of the 8 items measuring negative coping as the (baseline) covariate $L_1$ that potentially modified the (in)direct effects, with higher scores indicating greater tendency to use negative coping. To ease interpretation of the indirect effects, we standardized the sum scores so that the sample mean and standard deviation of negative coping were zero and one respectively.
The remaining observed covariates were age, gender, marital status, education level, working experience (in years) and job rank seniority; we denoted these variables, as well as negative coping, jointly by $L$. 
We assumed that the observed covariates $L$ were unaffected by exposure, mediators or outcome, and that they were sufficient for the identification assumptions \eqref{eq:identify_1}--\eqref{eq:identify_3} to hold. 
{\blue A causal diagram \citep{pearl2009} representing the causal relations between the variables, and the identification assumptions, is shown in Figure~\ref{fig:data}. We emphasize that the mediator indices are arbitrary, and do not reflect any a priori causal ordering among the mediators.}

\subsection{Analysis without effect modification}

\begin{figure}[!ht]
\centering
\caption{Weights used in the inverse weighting procedure to estimate the interventional (in)direct effects in the applied example. No mediator-covariate interaction terms were assumed in the mediator and outcome models. The row numbers index the duplicated data {\blue corresponding to Table~\ref{table:duplicatedall_inverseweighting} (but with an additional row because there were three mediators)} used to fit the interventional effect model. The weights were log-transformed (base 10), and only the non-zero weights for each row were then plotted. \label{fig:applied_ex:weights:noC}}
\includegraphics[width=\textwidth]{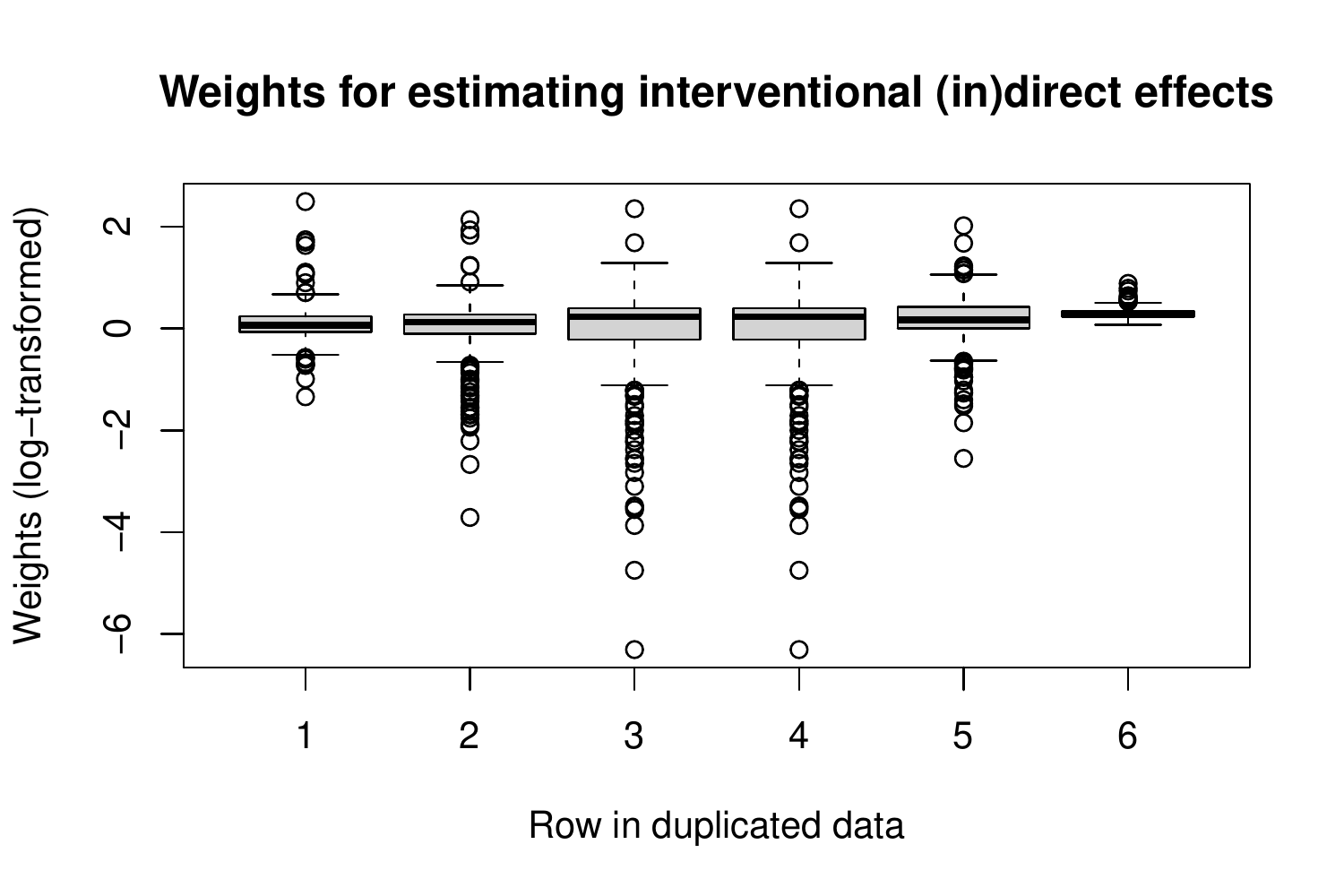}
\end{figure}

We first posited the following effect model for the interventional direct and indirect effects without any effect modification by the covariates.
\begin{align}
&\logit\left\{\EC\left(Y_{a m_1 m_2 m_3}\right)\right\} \notag\\
&= \mu_{0} + \left(\sum_{s=1}^{3} \theta_{s}a^{(s)}\right)(1-J) + 
\left\{ \mu_{0J} + \gamma_0 a^{(0)} + \gamma_1 a^{(1)}\indicator{a^{(1)}=a^{(2)}=a^{(3)}} \right\}J \notag\\
&\quad
+ \mu_{age} \hbox{age}
+ \mu_{gen} \hbox{gender} 
+ \mu_{mar} \hbox{marital status}
+ \mu_{edu} \hbox{education level} \notag\\
&\quad
+ \mu_{wor} \hbox{working experience}
+ \mu_{sen} \hbox{rank seniority}
+ \mu_{neg} \hbox{negative coping}.
\label{eq:effects_model_appliedex:noC}
\end{align}
The interventional indirect effects via each mediator $M_s, s=1,2,3$, were encoded by $\theta_s$. The joint indirect effect was encoded by $\gamma_{1}$, and the indirect effect via the mediators' mutual dependence was encoded by $\gamma_{1}-\sum_{s=1}^{3} \theta_{s}$. The direct effect was encoded by $\gamma_{0}$.
Because the exposure was not randomly assigned, we assumed the following propensity score model with main effects for all covariates in $L$ for the inverse weighting procedure.
\begin{align*}
&\logit\left\{\Pr(A=1|L)\right\} \\
&= 
\nu_0 + \nu_{age} \hbox{age}
+ \nu_{gen} \hbox{gender} 
+ \nu_{mar} \hbox{marital status}
+ \nu_{edu} \hbox{education level} \\
&\quad
+ \nu_{wor} \hbox{working experience}
+ \nu_{job} \hbox{rank seniority} 
+ \nu_{neg} \hbox{negative coping}.
\end{align*}
The joint mediator distribution was factorized as $f^a(\bm M|L) = f^a(M_1|L)f^a(M_2|M_1,L)f^a(M_3|M_1,M_2,L)$. 
Again we reiterate that the factorization does not assume or imply any causal ordering among the mediators.
We assumed all three mediators to be normally distributed, and fitted to the observed data the following linear regression models for the mediators within each exposure group $A=a$:
\begin{align*}
&\E(M_s | A=a, L) \\
&= \alpha_{s0}^a 
+ \alpha_{s,age}^a \hbox{age}
+ \alpha_{s,gen}^a \hbox{gender} 
+ \alpha_{s,mar}^a \hbox{marital status}
+ \alpha_{s,edu}^a \hbox{education level}\\
&\quad
+ \alpha_{s,wor}^a \hbox{working experience}
+ \alpha_{s,job}^a \hbox{rank seniority}
+ \alpha_{s,neg}^a \hbox{negative coping} 
, \quad s=1; \\
&\E(M_s | A=a, M_1, \ldots, M_{s-1}, L) \\
&= \eta_{s0}^a + \sum_{k=1}^{s-1} \eta_{s,k}^a M_k
+ \sum_{k,l=1: k>l}^{s-1} \eta_{s,k,l}^a M_kM_l \\
&\quad
+ \eta_{s,age}^a \hbox{age}
+ \eta_{s,gen}^a \hbox{gender} 
+ \eta_{s,mar}^a \hbox{marital status}
+ \eta_{s,edu}^a \hbox{education level}\\
&\quad
+ \eta_{s,wor}^a \hbox{working experience}
+ \eta_{s,job}^a \hbox{rank seniority}
+ \eta_{s,neg}^a \hbox{negative coping}
, \quad s=2,3.
\end{align*}
Constant error variances were assumed in each mediator model.
The Monte Carlo estimation procedure required further assuming the following logistic regression outcome model that was fitted within each exposure group $A=a$:
\begin{align*}
&\logit\{\Pr(Y=1|A=a,M_1,M_2,M_3,L)\} \\
&= \beta_0^a + \sum_{s=1}^{3} \beta_s^a M_s
+ \sum_{s,k=1:s>k}^{3} \beta_{s,k}^a M_sM_k \\
&\quad
+ \beta_{age}^a \hbox{age}
+ \beta_{gen}^a \hbox{gender}
+ \beta_{mar}^a \hbox{marital status} 
+ \beta_{edu}^a \hbox{education level}  \\
&\quad
+ \beta_{wor}^a \hbox{working experience}
+ \beta_{job}^a \hbox{rank seniority}
+ \beta_{neg}^a \hbox{negative coping}.
\end{align*}
{\blue Mediator-mediator interaction terms were included in the mediator and outcome models to allow the effect of each mediator to differ based on the levels of another mediator.}
We carried out both inverse weighting and Monte Carlo estimation procedures. 
The resulting weights using the former procedure were relatively unstable as shown in Figure~\ref{fig:applied_ex:weights:noC}.
{\blue Because there were three mediators in the applied example, the duplicated data for each individual corresponding to Table~\ref{table:duplicatedall_inverseweighting} consisted of six rows (instead of just five when there were only two mediators). The first four rows corresponded to the mediators under their respective marginal distributions, whereas the last two rows corresponded to the mediators under their joint distribution.}

{\blue Nonparametric percentile bootstrap 95\% confidence intervals were constructed using 10000 bootstrap samples that randomly resampled $n$ observations with replacement and repeated the estimation procedures for each bootstrap sample.}
Because the estimated interventional (in)direct effects using the inverse weighting procedure differed from those using the Monte Carlo procedure, and the weights appeared to be highly variable, we only presented the 95\% confidence intervals (CIs) for the Monte Carlo procedure in Table~\ref{table:appliedex_results-perm1:noC}. 
The estimated interventional indirect effects of self-efficacy on fatigue that were mediated through each of the subscales for the self-reported PTSD symptoms ($\theta_s, s=1,2,3$) suggested that shifting the (counterfactual) distribution of each subscale from those with self-efficacy levels above the empirical median to those with levels below the median resulted in lower fatigue on average, while holding the distributions of the other subscales fixed and setting self-efficacy to below the empirical median.
In particular, the effect of self-efficacy on fatigue that was mediated through the hyperarousal subscale ($\theta_3$) was estimated to be {\blue -0.30, with a 95\% CI of (-0.54, -0.03)}.
The interventional direct effect of self-efficacy on fatigue that was unmediated through the measured PTSD symptoms ($\gamma_0$) had the largest magnitude among the estimated (in)direct effects, with an estimated value of {\blue -0.76 (95\% CI=(-1.17, -0.44))}.

\begin{table}[ht]
\centering
\caption{Estimates and 95\% bootstrap (percentile) confidence intervals (``95\% CI'') for the interventional (in)direct effects, as encoded by the parameters in \eqref{eq:effects_model_appliedex:noC}, for the applied example. The estimates were obtained using either the inverse weighting (``IW'') procedure, or the imputation-based Monte Carlo (``MC'') procedure.
The CIs were calculated for the MC estimator only. 
No mediator-covariate interaction terms were assumed in the mediator and outcome models. 
All results were rounded to two decimal places.
\label{table:appliedex_results-perm1:noC}}
\begin{tabular}{lcrrc}
  \hline
Effect & Parameter & IW & MC & 95\% CI {\blue for MC} \\ 
  \hline
Indirect via re-experiencing & $\theta_{1}$ & 0.05 & -0.01 & (-0.21, 0.17) \\ 
Indirect via avoidance & $\theta_{2}$ & -0.53 & -0.22 & (-0.50, 0.07) \\ 
Indirect via hyperarousal &  $\theta_{3}$ & 0.00 & -0.30 & (-0.54, -0.03) \\ 
Indirect via mutual dependence of all mediators  & $\gamma_{1}-\sum_{s=1}^{3} \theta_{s}$ 
& -0.31 & -0.04 & (-0.25, 0.14) \\ 
Direct &  $\gamma_{0}$ & -0.56 & -0.76 & (-1.17, -0.44) \\ 
   \hline
\end{tabular}
\end{table}

The interventional indirect effects via each mediator as defined in this article can differ depending on the (arbitrary) indices used merely to label the mediators, because the other mediators are fixed at different hypothetical exposure levels. {\blue For example, when re-experiencing is indexed as $M_1$, the indirect effect via re-experiencing compares the (potential outcomes in the) first two rows of Tables~\ref{table:duplicatedall_inverseweighting} and \ref{table:duplicatedall}. But had re-experiencing been merely indexed as $M_2$ instead, the indirect effect would then correspond to the second and third rows.}
Nonetheless, we reiterate that the conceptual interpretations of the interventional indirect effect via each mediator are invariant to the choice of indices. We carried out a sensitivity analysis, by considering each of the $3!=6$ possible permutations of the three mediators in turn, and calculated the indirect effects (and the 95\% CIs) for each permutation using the Monte Carlo procedure. 
{\blue Each permutation of the indices implies a different decomposition of the joint indirect effect, and subsequently, may result in different estimates of the separate indirect effects. In practice, the substantive research question may be used to determine the most relevant decomposition (of the joint indirect effect) from a policy perspective, which subsequently implies the ordering of the mediator indices \citep{moreno2019defining}.}
The minimum and maximum estimates (and bounds of the 95\% CIs) across all the permutations are shown in Table~\ref{table:appliedex_results-permuted:noC}. The conclusions for the indirect effects were generally unchanged across the different decompositions resulting from different permutations of the mediator indices. 
Among the three PTSD subscales, the estimated interventional indirect effect was strongest for hyperarousal.

\begin{table}
\caption{Interventional indirect effect estimates and 95\% bootstrap (percentile) confidence intervals (``95\% CI'') using the Monte Carlo procedure for the applied example. The minimum (``min.'') and maximum (``max.'') estimates, and 95\% CI lower and upper bounds, across all $3!=6$ possible permutations of the mediator indices are presented. 
No mediator-covariate interaction terms were assumed in the mediator and outcome models. 
All results were rounded to two decimal places.
\label{table:appliedex_results-permuted:noC}}
\begin{center}
\begin{tabular}{lrrrrrr}
  \hline
 & \multicolumn{2}{c}{Estimate} & \multicolumn{2}{c}{95\% CI (lower)} & \multicolumn{2}{c}{95\% CI (upper)} \\
 Interventional indirect effect  & Min. & Max. &  Min. & Max. &  Min. & Max. \\ 
  \hline
 Re-experiencing & -0.10 & 0.01 & -0.32 & -0.21 & 0.15 & 0.24 \\
 Avoidance & -0.22 & -0.12 & -0.50 & -0.46 & 0.07 & 0.19 \\
 Hyperarousal & -0.32 & -0.30 & -0.56 & -0.54 & -0.07 & 0.00 \\
   \hline 
\end{tabular}
\end{center}
\end{table}

\subsection{Analysis with effect modification}

We further assessed whether the interventional indirect effects were modified by the baseline covariate negative coping, by positing the following effect model:
\begin{align}
&\logit\left\{\EC\left(Y_{a m_1 m_2 m_3}\right)\right\} \notag\\
&= \mu_{0} + 
\left\{ \sum_{s=1}^{3} \theta_{s}a^{(s)} + \theta_{sc}a^{(s)}\hbox{negative coping}\right\}(1-J)\notag\\ 
&\quad+ \left\{ \mu_{0J} + \gamma_0 a^{(0)} + \gamma_{0c} a^{(0)}\hbox{negative coping} \right. \notag\\
&\qquad+ \left. \gamma_1 a^{(1)}\indicator{a^{(1)}=a^{(2)}=a^{(3)}} + \gamma_{1c} a^{(1)}\indicator{a^{(1)}=a^{(2)}=a^{(3)}}\hbox{negative coping} \right\}J \notag\\
&\quad+ \mu_{age} \hbox{age}
+ \mu_{gen} \hbox{gender} 
+ \mu_{mar} \hbox{marital status}
+ \mu_{edu} \hbox{education level} \notag\\
&\quad
+ \mu_{wor} \hbox{working experience}
+ \mu_{sen} \hbox{rank seniority}
+ \mu_{neg} \hbox{negative coping}.
\label{eq:effects_model_appliedex}
\end{align}
The interventional indirect effects via each mediator $M_s, s=1,2,3$ (at the sample average level of negative coping) were encoded by $\theta_s$, and the effects were modified by negative coping when the parameters $\theta_{sc}$ were non-zero. Similarly, the joint indirect and direct effects were encoded by $\gamma_{1}$ and $\gamma_{0}$ respectively; the effect measures were modified by negative coping when $\gamma_{1c}$ and $\gamma_{0c}$ were non-zero respectively.

We carried out both inverse weighting and Monte Carlo estimation procedures. 
The inverse weighting estimators were calculated using the same fitted propensity score model as before.
The linear mediator mean models, and logistic regression outcome model, were augmented with interaction terms between the mediator(s) and negative coping to allow the effects of the mediators on each other, and on the outcome, to differ based on negative coping; i.e.,
\begin{align*}
&\E(M_s | A=a, M_1, \ldots, M_{s-1}, L) \\
&= \eta_{s0}^a + \sum_{k=1}^{s-1} (\eta_{s,k}^a + \eta_{s,k,neg}^a \hbox{negative coping}) M_k
+ \sum_{k,l=1: k>l}^{s-1} (\eta_{s,k,l}^a + \eta_{s,k,l,neg}^a \hbox{negative coping}) M_kM_l \\
&\quad
+ \eta_{s,age}^a \hbox{age}
+ \eta_{s,gen}^a \hbox{gender} 
+ \eta_{s,mar}^a \hbox{marital status}
+ \eta_{s,edu}^a \hbox{education level}\\
&\quad
+ \eta_{s,wor}^a \hbox{working experience}
+ \eta_{s,job}^a \hbox{rank seniority}
+ \eta_{s,neg}^a \hbox{negative coping}, \quad s=2,3; \\
&\logit\{\Pr(Y=1|A=a,M_1,M_2,M_3,L)\} \\
&= \beta_0^a + \sum_{s=1}^{3} (\beta_s^a + \beta_{s,neg}^a \hbox{negative coping}) M_s
+ \sum_{s,k=1:s>k}^{3} (\beta_{s,k}^a + \beta_{s,k,neg}^a \hbox{negative coping}) M_sM_k \\
&\quad
+ \beta_{age}^a \hbox{age}
+ \beta_{gen}^a \hbox{gender}
+ \beta_{mar}^a \hbox{marital status} 
+ \beta_{edu}^a \hbox{education level}  \\
&\quad
+ \beta_{wor}^a \hbox{working experience}
+ \beta_{job}^a \hbox{rank seniority}
+ \beta_{neg}^a \hbox{negative coping}.
\end{align*}
The results from fitting the models to the observed data are displayed in Appendix~\ref{sect:appliedex:regcoef}.
Because the estimated weights for the inverse weighting estimator were similarly unstable as those shown in Figure~\ref{fig:applied_ex:weights:noC}, we again only calculated the 95\% CIs for the Monte Carlo estimator. 
The results displayed in Table~\ref{table:appliedex_results-perm1} suggested that negative coping primarily modified the indirect effect via the hyperarousal subscale for the self-reported PTSD symptoms ($\theta_{3c}$), with the estimate of $\theta_{3c}$ having the largest (absolute) magnitude among the three subscales $\theta_{sc}, s=1,2,3$.
The estimated reduction in fatigue - due to the indirect effect of self-efficacy via hyperarousal - tended to be stronger for those with higher negative coping levels; e.g., at one standard deviation above the sample mean of negative coping, there was an estimated further reduction of {\blue -0.19 (95\% CI=(-0.42, 0.15))}.
Conversely, the effect of self-efficacy that was mediated through the re-experiencing subscale was higher fatigue among those with higher negative coping levels; e.g., the estimated increase in fatigue at one standard deviation above the sample mean of negative coping was {\blue 0.14 (95\% CI=(-0.20, 0.33))}.
Similar conclusions from the previous analysis were arrived at. The estimated interventional indirect effects of self-efficacy on fatigue were primarily mediated through the hyperarousal subscale. The interventional direct effect that was unmediated through the measured PTSD symptoms ($\gamma_0$) had the largest magnitude among the estimated (in)direct effects, with an estimated value of {\blue -0.76 (95\% CI=(-1,19, -0.44))} at the sample mean of negative coping.
The conclusions for the indirect effects were generally unchanged after considering the different decompositions in a sensitivity analysis, as shown in Table~\ref{table:appliedex_results-permuted}.

\begin{table}[ht]
\centering
\caption{Estimates and 95\% bootstrap (percentile) confidence intervals (``95\% CI'') for the interventional (in)direct effects, as encoded by the parameters in \eqref{eq:effects_model_appliedex}, for the applied example. The estimates were obtained using either the inverse weighting (``IW'') procedure, or the imputation-based Monte Carlo (``MC'') procedure.
The CIs were calculated for the MC estimator only.
All results were rounded to two decimal places.
\label{table:appliedex_results-perm1}}
\begin{tabular}{lrrrrrr}
  \hline
Effect & Parameter & IW & MC & \multicolumn{2}{c}{95\% CI} \\ 
  \hline
Indirect via re-experiencing & $\theta_{1}$ & 0.01 & 0.02 & -0.22 & 0.18 \\ 
Indirect via re-experiencing (with negative coping) & $\theta_{1c}$ & -0.11 & 0.14 & -0.20 & 0.33 \\ 
Indirect via avoidance & $\theta_{2}$ & -0.55 & -0.23 & -0.51 & 0.11 \\ 
Indirect via avoidance (with negative coping) &  $\theta_{2c}$ & -0.22 & -0.01 & -0.40 & 0.34 \\ 
Indirect via hyperarousal &  $\theta_{3}$ & 0.00 & -0.33 & -0.56 & -0.02 \\ 
Indirect via hyperarousal (with negative coping) &  $\theta_{3c}$ & -0.00 & -0.19 & -0.42 & 0.15 \\ 
Indirect via mutual dependence & $\gamma_{1}-\sum_{s=1}^{3} \theta_{s}$  & 
-0.16 & -0.00 & -0.27 & 0.16 \\ 
Indirect via mutual dependence (with negative coping)  & $\gamma_{1c}-\sum_{s=1}^{3} \theta_{sc}$  & -0.11 & -0.08 & -0.21 & 0.21 \\ 
Direct &  $\gamma_{0}$ & -0.63 & -0.76 & -1.19 & -0.44 \\ 
Direct (with negative coping) & $\gamma_{0c}$ & 0.50 & 0.07 & -0.33 & 0.42 \\ 
   \hline
\end{tabular}
\end{table}

\begin{table}
\caption{Interventional indirect effect estimates and 95\% bootstrap (percentile) confidence intervals (``CI'') using the Monte Carlo procedure for the applied example. The minimum (``min.'') and maximum (``max.'') estimates, and 95\% CI lower and upper bounds, across all $3!=6$ possible permutations of the mediator indices are presented. 
{\blue Indirect effects that were possibly modified by negative coping, and thus corresponded to the interaction terms in the effect model, were denoted using a colon (``:''). The remaining indirect effects corresponded to the main effects in the effect model.}
All results were rounded to two decimal places.
\label{table:appliedex_results-permuted}}
\begin{center}
\begin{tabular}{lrrrrrr}
  \hline
 & \multicolumn{2}{c}{Estimate} & \multicolumn{2}{c}{95\% CI (lower)} & \multicolumn{2}{c}{95\% CI (upper)} \\
 Interventional indirect effect  & Min. & Max. &  Min. & Max. &  Min. & Max. \\ 
  \hline
 Re-experiencing & -0.09 & 0.02 & -0.32 & -0.22 & 0.16 & 0.23 \\ 
 Re-experiencing:Negative coping & -0.02 & 0.14 & -0.25 & -0.20 & 0.25 & 0.33 \\   
 Avoidance & -0.23 & -0.13 & -0.51 & -0.47 & 0.11 & 0.19 \\ 
 Avoidance:Negative coping & -0.01 & 0.11 & -0.42 & -0.34 & 0.33 & 0.40 \\ 
 Hyperarousal & -0.33 & -0.30 & -0.57 & -0.55 & -0.04 & 0.04 \\ 
 Hyperarousal:Negative coping & -0.24 & -0.14 & -0.49 & -0.42 & 0.15 & 0.25 \\ 
   \hline       
\end{tabular}
\end{center}
\end{table}         

\section{Discussion \label{sect:discussion}}
In this article we introduced interventional effect models for estimating interventional indirect effects for multiple mediators that may be modified by observed baseline covariates. 
We described two estimation procedures for the causal parameters that index the effect models: inverse weighting by the counterfactual mediator density or mass functions, or Monte Carlo-based integration. 
The former approach may be better suited when the mediator weights are relatively stable, such as when the mediators are weakly (or un)affected by exposure or each other, whereas the latter approach may be better suited when it is reasonable to assume that the parametric outcome model is correct.

There are several avenues of possible future research related to mediation analyses with multiple mediators using interventional effects developed in this paper. 
While the interventional indirect effects defined in this paper have been fixed at the hypothetical exposure level $a^{(0)}=0$, in principle, the indirect effects can be defined using $a^{(0)}=1$. More general definitions of the interventional indirect effects permit exploring different decompositions, by including interactions between the hypothetical exposures, e.g., $a^{(0)}$ with either $a^{(1)}$ or $a^{(2)}$, or both, in the posited effect model. 
{\blue Similar to the natural effect models of \citet{lange2012simple} and \citet{steen2017flexible}, the interactions encode (in)direct effects that may be modified by the values of the mediators (and exposure) under different (hypothetical) exposure levels.
For example, augmenting the effect model in \eqref{eq:effects_model_appliedex:noC} with an additional term $\gamma_{01} a^{(0)}a^{(1)}\indicator{a^{(1)}=a^{(2)}=a^{(3)}}J$ permits investigating two different decompositions of the total effect (concurrently within the same effect model).
Under the first decomposition, the direct and (joint) indirect effects, as previously defined in \eqref{eq:DE_define} and \eqref{eq:IE_define}, are respectively encoded by:
$
\gamma_0 + \gamma_{01} = 
g\left\{\EC(Y_{1\tilde {\bm M}_{1}})\right\} - g\left\{\EC(Y_{0\tilde {\bm M}_{1}})\right\},
\gamma_1 = 
g\left\{\EC(Y_{0\tilde {\bm M}_{1}})\right\} - g\left\{\EC(Y_{0\tilde {\bm M}_{0}})\right\}.
$
Whereas under the second decomposition, the direct and (joint) indirect effects are respectively encoded by:
$
\gamma_0 = 
g\left\{\EC(Y_{1\tilde {\bm M}_{0}})\right\} - g\left\{\EC(Y_{0\tilde {\bm M}_{0}})\right\},
\gamma_1 + \gamma_{01} = 
g\left\{\EC(Y_{1\tilde {\bm M}_{1}})\right\} - g\left\{\EC(Y_{1\tilde {\bm M}_{0}})\right\},
$
so that the direct effect holds the joint distribution of the (counterfactual) mediators fixed under the (hypothetical) control level. Researchers may therefore use such interactions in an effect model to evaluate the practical relevance of different possible decompositions.}
However, more complex interventional effect models may be required, with additional interaction terms between the covariate(s) and the (mediators') exposure levels when the interventional effects may be modified by baseline covariates. Furthermore, estimation using the inverse weighting procedure may require calculating (non-zero) weights for all individuals that may exacerbate any finite sample biases due to unstable weights.
{\blue The investigation of weight stabilization strategies such as truncation \citep{imai2015robust}, binning \citep{naimi2014mediation}, or optimization-based algorithms \citep{santacatterina2019optimal}, to improve the inverse weighting estimator should thus be carried out in future work.

Effect heterogeneity in this paper is restricted to pre-specified statistical interactions between baseline covariates and hypothetical exposure levels in the effect model, and thus interaction effects between baseline covariates, and exposure or the mediators, or both, in the fitted mediator and outcome models. Using machine learning-based methods to perform variable selection for covariates that may modify the indirect effects, by extending \citet{imai2013estimating} who focus on a single exposure with no mediators, or adapting the debiasing algorithms for heterogeneous treatment effects as in \citet{nie2020quasioracle} or \citet{kennedy2020optimal}, are areas for future work. Nonparametric inference along the lines of the latter two works is additionally promising because it can theoretically eliminate any extrapolation bias arising from an outcome model that is uncongenial with the effect model; such nonparametric inference is achievable, even when parametric effect models are used to provide parsimonious effect summaries \citep{vansteelandt2020assumptionlean}. 
The proposed methods in this paper are intended to simplify estimation of interventional indirect effects, and is thus relevant to researchers exploring heterogeneous mediated effects via multiple possible mediators, without having to specify a (possibly arbitrary) causal structure among the mediators. When substantive interest is in finding a subset of the (possibly large number of) mediators for further investigation, a threshold that either is pre-determined, such as in \citet{huang2016hypothesis}, or controls the familywise error rate or the false discovery rate under multiple testing scenarios, such as in \citet{dai2020multiple} or \citet{derkach2020group}, may be employed.}

\section*{Acknowledgements}
This research was supported by the Research Foundation -- Flanders (FWO) under Grant G019317N. 
Computational resources and services were provided by the VSC (Flemish Supercomputer Center), funded by the FWO and the Flemish Government -- department EWI.
The content is solely the responsibility of the authors and does not 
represent the official views of the authors' institutions or FWO.

\bibliographystyle{plainnat}
\bibliography{../interventional}

\begin{thebibliography}{47}
\providecommand{\natexlab}[1]{#1}
\providecommand{\url}[1]{\texttt{#1}}
\expandafter\ifx\csname urlstyle\endcsname\relax
  \providecommand{\doi}[1]{doi: #1}\else
  \providecommand{\doi}{doi: \begingroup \urlstyle{rm}\Url}\fi

\bibitem[Albert et~al.(2019)Albert, Cho, Liu, and Nelson]{albert2019gmediation}
Jeffrey~M Albert, Jang~Ik Cho, Yiying Liu, and Suchitra Nelson.
\newblock Generalized causal mediation and path analysis: Extensions and
  practical considerations.
\newblock \emph{Statistical Methods in Medical Research}, 28\penalty0
  (6):\penalty0 1793--1807, 2019.
\newblock \doi{10.1177/0962280218776483}.

\bibitem[Andrews and Didelez(2020)]{andrews2020insights}
Ryan~M Andrews and Vanessa Didelez.
\newblock Insights into the ``cross-world'' independence assumption of causal
  mediation analysis.
\newblock \emph{arXiv preprint arXiv:2003.10341}, 2020.

\bibitem[Avin et~al.(2005)Avin, Shpitser, and Pearl]{avin2005identifiability}
Chen Avin, Ilya Shpitser, and Judea Pearl.
\newblock Identifiability of path-specific effects.
\newblock In \emph{Proceedings of the 19th International Joint Conference on
  Artificial Intelligence}, pages 357--363, San Francisco, CA, USA, July 2005.
  Morgan Kaufmann Publishers Inc.

\bibitem[Dai et~al.(2020)Dai, Stanford, and LeBlanc]{dai2020multiple}
James~Y. Dai, Janet~L. Stanford, and Michael LeBlanc.
\newblock A multiple-testing procedure for high-dimensional mediation
  hypotheses.
\newblock \emph{Journal of the American Statistical Association}, 0\penalty0
  (0):\penalty0 1--16, 2020.
\newblock \doi{10.1080/01621459.2020.1765785}.
\newblock URL \url{https://doi.org/10.1080/01621459.2020.1765785}.

\bibitem[Daniel et~al.(2015)Daniel, De~Stavola, Cousens, and
  Vansteelandt]{daniel2015causal}
Rhian~M Daniel, Bianca~L De~Stavola, S~N Cousens, and Stijn Vansteelandt.
\newblock Causal mediation analysis with multiple mediators.
\newblock \emph{Biometrics}, 71\penalty0 (1):\penalty0 1--14, 2015.
\newblock \doi{10.1111/biom.12248}.

\bibitem[Derkach et~al.(2020)Derkach, Moore, Boca, and
  Sampson]{derkach2020group}
Andriy Derkach, Steven~C Moore, Simina~M Boca, and Joshua~N Sampson.
\newblock Group testing in mediation analysis.
\newblock \emph{Statistics in Medicine}, 2020.

\bibitem[Didelez et~al.(2006)Didelez, Dawid, and Geneletti]{didelez2006direct}
Vanessa Didelez, A~Philip Dawid, and Sara Geneletti.
\newblock Direct and indirect effects of sequential treatments.
\newblock In \emph{Proceedings of the 22nd Conference on Uncertainty in
  Artificial Intelligence}, pages 138--146, Arlington, VA, USA, 2006. AUAI
  Press.

\bibitem[Efron and Tibshirani(1994)]{efron1994introduction}
Bradley Efron and Robert~J Tibshirani.
\newblock \emph{An introduction to the bootstrap}.
\newblock Chapman and Hall/CRC, 1994.
\newblock \doi{10.1007/978-1-4899-4541-9}.

\bibitem[Greenland et~al.(1999)Greenland, Robins, and
  Pearl]{greenland1999confounding}
Sander Greenland, James~M. Robins, and Judea Pearl.
\newblock Confounding and collapsibility in causal inference.
\newblock \emph{Statistical Science}, 14\penalty0 (1):\penalty0 29--46, 1999.

\bibitem[Hong(2010)]{hong2010ratio}
Guanglei Hong.
\newblock Ratio of mediator probability weighting for estimating natural direct
  and indirect effects.
\newblock In American~Statistical Association, editor, \emph{Proceedings of the
  American Statistical Association, Biometrics Section}, pages 2401--2415,
  Alexandria, VA, USA, 2010.

\bibitem[Hou et~al.(2020)Hou, Dong, Zhang, Song, Zhang, Cai, Liu, and
  Deng]{hou2020selfefficacy}
Tianya Hou, Wei Dong, Ruike Zhang, Xiangrui Song, Fan Zhang, Wenpeng Cai, Ying
  Liu, and Guanghui Deng.
\newblock Self-efficacy and fatigue among health care workers during covid-19
  outbreak: A moderated mediation model of posttraumatic stress disorder
  symptoms and negative coping.
\newblock \emph{PREPRINT (Version 1) available at Research Square}, 2020.
\newblock \doi{10.21203/rs.3.rs-23066/v1}.

\bibitem[Huang and Pan(2016)]{huang2016hypothesis}
Yen-Tsung Huang and Wen-Chi Pan.
\newblock Hypothesis test of mediation effect in causal mediation model with
  high-dimensional continuous mediators.
\newblock \emph{Biometrics}, 72\penalty0 (2):\penalty0 402--413, 2016.
\newblock \doi{10.1111/biom.12421}.

\bibitem[Imai and Ratkovic(2013)]{imai2013estimating}
Kosuke Imai and Marc Ratkovic.
\newblock Estimating treatment effect heterogeneity in randomized program
  evaluation.
\newblock \emph{The Annals of Applied Statistics}, 7\penalty0 (1):\penalty0
  443--470, 2013.

\bibitem[Imai and Ratkovic(2015)]{imai2015robust}
Kosuke Imai and Marc Ratkovic.
\newblock Robust estimation of inverse probability weights for marginal
  structural models.
\newblock \emph{Journal of the American Statistical Association}, 110\penalty0
  (511):\penalty0 1013--1023, 2015.

\bibitem[Jackson and VanderWeele(2018)]{jackson2018decomposition}
John~W Jackson and Tyler~J VanderWeele.
\newblock Decomposition analysis to identify intervention targets for reducing
  disparities.
\newblock \emph{Epidemiology}, 29\penalty0 (6):\penalty0 825--835, 2018.
\newblock \doi{10.1097/EDE.0000000000000901}.

\bibitem[Kennedy(2020)]{kennedy2020optimal}
Edward~H. Kennedy.
\newblock Optimal doubly robust estimation of heterogeneous causal effects.
\newblock 2020.

\bibitem[Lange et~al.(2012)Lange, Vansteelandt, and Bekaert]{lange2012simple}
Theis Lange, Stijn Vansteelandt, and Maarten Bekaert.
\newblock A simple unified approach for estimating natural direct and indirect
  effects.
\newblock \emph{American Journal of Epidemiology}, 176\penalty0 (3):\penalty0
  190--195, 2012.

\bibitem[Lange et~al.(2013)Lange, Rasmussen, and Thygesen]{lange2013assessing}
Theis Lange, Mette Rasmussen, and Lau~Caspar Thygesen.
\newblock Assessing natural direct and indirect effects through multiple
  pathways.
\newblock \emph{American Journal of Epidemiology}, 179\penalty0 (4):\penalty0
  513--518, 2013.
\newblock \doi{10.1093/aje/kwt270}.

\bibitem[Lin and VanderWeele(2017)]{lin2017interventional}
Sheng-Hsuan Lin and Tyler VanderWeele.
\newblock Interventional approach for path-specific effects.
\newblock \emph{Journal of Causal Inference}, 5\penalty0 (1), 2017.
\newblock \doi{10.1515/jci-2015-0027}.

\bibitem[Lok(2019)]{lok2019organic}
Judith~J Lok.
\newblock {Causal organic direct and indirect effects: closer to Baron and
  Kenny}.
\newblock \emph{arXiv preprint arXiv:1903.04697}, art. arXiv:1903.04697, Mar
  2019.

\bibitem[Meng(1994)]{meng1994multiple}
Xiao-Li Meng.
\newblock Multiple-imputation inferences with uncongenial sources of input.
\newblock \emph{Statistical Science}, 9\penalty0 (4):\penalty0 538--558, 1994.

\bibitem[Micali et~al.(2018)Micali, Daniel, Ploubidis, and
  De~Stavola]{micali2018maternal}
Nadia Micali, Rhian~M Daniel, George~B Ploubidis, and Bianca~L De~Stavola.
\newblock Maternal prepregnancy weight status and adolescent eating disorder
  behaviors: A longitudinal study of risk pathways.
\newblock \emph{Epidemiology}, 29\penalty0 (4):\penalty0 579--589, 2018.

\bibitem[Moreno-Betancur and Carlin(2018)]{moreno2018understanding}
Margarita Moreno-Betancur and John~B Carlin.
\newblock Understanding interventional effects: A more natural approach to
  mediation analysis?
\newblock \emph{Epidemiology}, 29\penalty0 (5):\penalty0 614--617, 2018.
\newblock \doi{10.1097/EDE.0000000000000866}.

\bibitem[{Moreno-Betancur} et~al.(2020){Moreno-Betancur}, {Moran}, {Becker},
  {Patton}, and {Carlin}]{moreno2019defining}
Margarita {Moreno-Betancur}, Paul {Moran}, Denise {Becker}, George {Patton},
  and John~B {Carlin}.
\newblock Mediation effects that emulate a target randomised trial:
  Simulation-based evaluation of ill-defined interventions on multiple
  mediators.
\newblock \emph{arXiv e-prints}, art. arXiv:1907.06734, July 2020.

\bibitem[Naimi et~al.(2014)Naimi, Kaufman, and MacLehose]{naimi2014mediation}
Ashley~I Naimi, Jay~S Kaufman, and Richard~F MacLehose.
\newblock Mediation misgivings: ambiguous clinical and public health
  interpretations of natural direct and indirect effects.
\newblock \emph{International Journal of Epidemiology}, 43\penalty0
  (5):\penalty0 1656--1661, 2014.

\bibitem[Nie and Wager(2020)]{nie2020quasioracle}
Xinkun Nie and Stefan Wager.
\newblock Quasi-oracle estimation of heterogeneous treatment effects.
\newblock 2020.

\bibitem[Noguchi(2020)]{npr2020pandemic}
Yuki Noguchi.
\newblock Pandemic affects mental health of frontline health workers, April
  2020.
\newblock URL
  \url{https://www.npr.org/2020/04/22/841925658/pandemic-affects-mental-health-of-frontline-health-workers}.

\bibitem[Pearl(2001)]{pearl2001direct}
Judea Pearl.
\newblock Direct and indirect effects.
\newblock In \emph{Proceedings of the 17th Conference on Uncertainty in
  Artificial Intelligence}, pages 411--420, San Francisco, CA, USA, 2001.
  Morgan Kaufmann Publishers Inc.

\bibitem[Pearl(2009)]{pearl2009}
Judea Pearl.
\newblock \emph{Causality: Models, Reasoning and Inference}.
\newblock Cambridge University Press, New York, NY, USA, 2nd edition, 2009.
\newblock \doi{10.1017/CBO9780511803161}.

\bibitem[Petersen et~al.(2006)Petersen, Sinisi, and van~der
  Laan]{petersen2006estimation}
Maya~L Petersen, Sandra~E Sinisi, and Mark~J van~der Laan.
\newblock Estimation of direct causal effects.
\newblock \emph{Epidemiology}, 17\penalty0 (3):\penalty0 276--284, 2006.
\newblock \doi{10.1097/01.ede.0000208475.99429.2d}.

\bibitem[Quynh~Nguyen et~al.(2019)Quynh~Nguyen, Schmid, and
  Stuart]{trang2019clarifying}
Trang Quynh~Nguyen, Ian Schmid, and Elizabeth~A. Stuart.
\newblock Clarifying causal mediation analysis for the applied researcher:
  {D}efining effects based on what we want to learn.
\newblock \emph{arXiv preprint arXiv:1904.08515}, art. arXiv:1904.08515, Apr
  2019.

\bibitem[Robins(2000)]{robins2000marginal}
James~M Robins.
\newblock \emph{Marginal structural models versus structural nested models as
  tools for causal inference}, pages 95--133.
\newblock Springer-Verlag, New York, NY, USA, 2000.
\newblock \doi{10.1007/978-1-4612-1284-3_2}.

\bibitem[Robins and Greenland(1992)]{robins1992identifiability}
James~M Robins and Sander Greenland.
\newblock Identifiability and exchangeability for direct and indirect effects.
\newblock \emph{Epidemiology}, 3\penalty0 (2):\penalty0 143--155, 1992.
\newblock \doi{10.1097/00001648-199203000-00013}.

\bibitem[Robins and Richardson(2010)]{robins2010alternative}
James~M Robins and Thomas~S Richardson.
\newblock \emph{Alternative graphical causal models and the identification of
  direct effects}, pages 103--158.
\newblock Oxford University Press, New York, NY, USA, 2010.
\newblock ISBN 9780199754649.

\bibitem[Santacatterina et~al.(2019)Santacatterina, Garc{\'\i}a-Pareja,
  Bellocco, S{\"o}nnerborg, Ekstr{\"o}m, and Bottai]{santacatterina2019optimal}
Michele Santacatterina, Celia Garc{\'\i}a-Pareja, Rino Bellocco, Anders
  S{\"o}nnerborg, Anna~Mia Ekstr{\"o}m, and Matteo Bottai.
\newblock Optimal probability weights for estimating causal effects of
  time-varying treatments with marginal structural cox models.
\newblock \emph{Statistics in Medicine}, 38\penalty0 (10):\penalty0 1891--1902,
  2019.
\newblock \doi{10.1002/sim.8080}.
\newblock URL \url{https://onlinelibrary.wiley.com/doi/abs/10.1002/sim.8080}.

\bibitem[Snowden et~al.(2011)Snowden, Rose, and
  Mortimer]{snowden2011implementation}
Jonathan~M Snowden, Sherri Rose, and Kathleen~M Mortimer.
\newblock Implementation of {G}-computation on a simulated data set:
  demonstration of a causal inference technique.
\newblock \emph{American Journal of Epidemiology}, 173\penalty0 (7):\penalty0
  731--738, 2011.

\bibitem[Steen et~al.(2017)Steen, Loeys, Moerkerke, and
  Vansteelandt]{steen2017flexible}
Johan Steen, Tom Loeys, Beatrijs Moerkerke, and Stijn Vansteelandt.
\newblock Flexible mediation analysis with multiple mediators.
\newblock \emph{American Journal of Epidemiology}, 186\penalty0 (2):\penalty0
  184--193, 2017.
\newblock \doi{10.1093/aje/kwx051}.

\bibitem[Taguri et~al.(2018)Taguri, Featherstone, and Cheng]{taguri2018}
Masataka Taguri, John Featherstone, and Jing Cheng.
\newblock Causal mediation analysis with multiple causally non-ordered
  mediators.
\newblock \emph{Statistical Methods in Medical Research}, 27\penalty0
  (1):\penalty0 3--19, 2018.
\newblock \doi{10.1177/0962280215615899}.

\bibitem[Tchetgen~Tchetgen(2014)]{tchetgen2014note}
Eric~J Tchetgen~Tchetgen.
\newblock A note on formulae for causal mediation analysis in an odds ratio
  context.
\newblock \emph{Epidemiologic Methods}, 2\penalty0 (1):\penalty0 21--31, 2014.

\bibitem[VanderWeele and Tchetgen~Tchetgen(2017)]{vanderweele2017mediation}
Tyler~J VanderWeele and Eric~J Tchetgen~Tchetgen.
\newblock Mediation analysis with time varying exposures and mediators.
\newblock \emph{Journal of the Royal Statistical Society: Series B (Statistical
  Methodology)}, 79\penalty0 (3):\penalty0 917--938, 2017.
\newblock \doi{10.1111/rssb.12194}.

\bibitem[VanderWeele and Vansteelandt(2010)]{vanderweele2010odds}
Tyler~J VanderWeele and Stijn Vansteelandt.
\newblock Odds ratios for mediation analysis for a dichotomous outcome.
\newblock \emph{American Journal of Epidemiology}, 172\penalty0 (12):\penalty0
  1339--1348, 2010.

\bibitem[VanderWeele et~al.(2014)VanderWeele, Vansteelandt, and
  Robins]{vanderweele2014effect}
Tyler~J VanderWeele, Stijn Vansteelandt, and James~M Robins.
\newblock Effect decomposition in the presence of an exposure-induced
  mediator-outcome confounder.
\newblock \emph{Epidemiology}, 25\penalty0 (2):\penalty0 300, 2014.
\newblock \doi{10.1097/EDE.0000000000000034}.

\bibitem[Vansteelandt and Daniel(2017)]{vansteelandt2017interventional}
Stijn Vansteelandt and Rhian~M. Daniel.
\newblock Interventional effects for mediation analysis with multiple
  mediators.
\newblock \emph{Epidemiology}, 28\penalty0 (2):\penalty0 258--265, 2017.
\newblock \doi{10.1097/EDE.0000000000000596}.

\bibitem[Vansteelandt and Dukes(2020)]{vansteelandt2020assumptionlean}
Stijn Vansteelandt and Oliver Dukes.
\newblock Assumption-lean inference for generalised linear model parameters.
\newblock 2020.

\bibitem[Vansteelandt and Keiding(2011)]{vansteelandt2011invited}
Stijn Vansteelandt and Niels Keiding.
\newblock Invited commentary: {G}-computation--lost in translation?
\newblock \emph{American Journal of Epidemiology}, 173\penalty0 (7):\penalty0
  739--742, 2011.

\bibitem[Vansteelandt and VanderWeele(2012)]{vansteelandt2012natural}
Stijn Vansteelandt and Tyler~J VanderWeele.
\newblock Natural direct and indirect effects on the exposed: effect
  decomposition under weaker assumptions.
\newblock \emph{Biometrics}, 68\penalty0 (4):\penalty0 1019--1027, 2012.

\bibitem[Vansteelandt et~al.(2012)Vansteelandt, Bekaert, and
  Lange]{vansteelandt2012imputation}
Stijn Vansteelandt, Maarten Bekaert, and Theis Lange.
\newblock Imputation strategies for the estimation of natural direct and
  indirect effects.
\newblock \emph{Epidemiologic Methods}, 1\penalty0 (1):\penalty0 131--158,
  2012.

\end{thebibliography}

\clearpage

\bigskip
\begin{center}
{\large\bf SUPPLEMENTARY MATERIAL}
\end{center}

\appendix

\section{Validity of the proposed procedure \label{sect:consistency}}

{\blue In this section, we derive the consistency of the Monte Carlo-based and inverse weighting estimators under the identifying assumptions \eqref{eq:identify_1}--\eqref{eq:identify_3}, and the positivity and (causal) consistency assumptions, described in the main text.} The estimands are identified by the observed quantities:
\begin{align*}
&\E(Y_{a^{(0)}\tilde {\bm M}_{a^{(1)}}}|L) \\
&= \int \E(Y_{a^{(0)} m_1\cdots m_t}|L) \; 
dF(M_{1,a^{(1)}}\!=\!m_1,\ldots,M_{t,a^{(1)}}\!=\!m_t| L) \\
&= \int \E\left(Y_{a^{(0)} m_1 \cdots m_t}|A=a^{(0)}, M_1=m_1, \ldots, M_t=m_t, L\right) dF(M_{1,a^{(1)}}\!=\!m_1,\ldots,M_{t,a^{(1)}}\!=\!m_t| A=a^{(1)}, L) \\
&= \int \E\left(Y |A=a^{(0)}, M_1=m_1, \ldots, M_t=m_t, L\right) 
dF(M_{1}\!=\!m_1,\ldots,M_{t}\!=\!m_t| A=a^{(1)}, L); \\
&\E(Y_{a^{(0)}\tilde M_{1,a^{(1)}}\cdots\tilde M_{t,a^{(t)}}}|L) \\
&=\int \E(Y_{a^{(0)} m_1\cdots m_t}|L) \; 
dF(M_{1,a^{(1)}}\!=\!m_1| L) \cdots dF(M_{t,a^{(t)}}\!=\!m_t| L) \\
&= \int \E\left(Y_{a^{(0)} m_1 \cdots m_t}|A=a^{(0)}, M_1=m_1, \ldots, M_t=m_t, L\right) \\
&\quad\quad\quad dF(M_{1,a^{(1)}}\!=\!m_1| A=a^{(1)}, L) \cdots dF(M_{t,a^{(t)}}\!=\!m_t| A=a^{(t)}, L) \\
&= \int \E\left(Y |A=a^{(0)}, M_1=m_1, \ldots, M_t=m_t, L\right) 
dF(M_{1}\!=\!m_1| A=a^{(1)}, L) \cdots dF(M_{t}\!=\!m_t| A=a^{(t)}, L).
\end{align*}
Unbiased estimation of the interventional effects using the Monte Carlo-based procedure therefore depends on correctly specifying (i) an outcome model conditional on exposure, mediators, and covariates that is unbiased for the marginal structural mean model, i.e., 
$\E\left(Y|a, m_1, \ldots, m_t, L\right)$ converges in probability to $\E\left(Y_{a m_1\cdots m_t}|L\right)$; and (ii) models for the (marginal) distributions of the mediators, conditional on exposure and covariates, that are unbiased for their counterfactual distributions, i.e., for $s=1,\ldots,t$, $f(M_{s} |A=a, L)$ converges in probability to its counterfactual distributions $f(M_{s,a} |L)$, for $a=0,1$. 

The form of the weights for the inverse weighting estimator suggested in the main text are motivated by further deriving the estimands as:
\begin{align*}
\E(Y_{a^{(0)}\tilde {\bm M}_{a^{(1)}}}|L)
&= \int \E\left(Y |A=a^{(0)}, M_1=m_1, \ldots, M_t=m_t, L\right) 
dF(\bm M = \bm m| A=a^{(1)}, L) \\
&= \int \int y f_Y\left(y |A=a^{(0)}, M_1=m_1, \ldots, M_t=m_t, L\right) 
f(\bm M = \bm m |A=a^{(1)}, L) \; dy \, d\bm m \\
&= \int \int y \dfrac{f\left(y, A=a^{(0)}, M_1=m_1, \ldots, M_t=m_t | L\right)}{f\left(A=a^{(0)}, M_1=m_1, \ldots, M_t=m_t | L\right)}f(\bm M = \bm m |A=a^{(1)}, L) \; dy \, d\bm m \\
&= \int\int y \indicator{A=a^{(0)}} f\left(y, A=a^{(0)}, M_1=m_1, \ldots, M_t=m_t | L\right) \\
& \quad\quad\quad \times
\underbrace{\dfrac{1}{\Pr(A=a^{(0)} | L)}}_{\mbox{$=w^a$}} \times
\underbrace{\dfrac{f(\bm M = \bm m |A=a^{(1)}, L)}{f(\bm M = \bm m | A=a^{(0)},L)}}_{\mbox{$=w^m(a^{(1)},\ldots,a^{(1)},J=1)$}} 
\; dy \, d\bm m \\
&= \E\left\{Y\indicator{A=a^{(0)}}w(a^{(0)},\ldots,a^{(1)},J=1) | L \right\},
\end{align*}
where $\bm m = (m_1, \ldots, m_t)$ for notational simplicity.
The estimands where the (counterfactual) mediators are randomly sampled from their marginal distributions (unconditional on any other mediators) are similarly derived as:
\begin{align*}
\E(Y_{a^{(0)}\tilde M_{1,a^{(1)}}\cdots\tilde M_{t,a^{(t)}}}|L)
&= \int\int y \indicator{A=a^{(0)}} f\left(y, A=a^{(0)}, M_1=m_1, \ldots, M_t=m_t | L\right) \\
& \quad\quad\quad \times
\underbrace{\dfrac{1}{\Pr(A=a^{(0)} | L)}}_{\mbox{$=w^a$}} \times
\underbrace{\dfrac{\prod_{s=1}^{t} f(M_s = m_s |A=a^{(s)}, L)}{f(\bm M = \bm m|A=a^{(0)}, L)}}_{\mbox{$=w^m(a^{(1)},\ldots,a^{(t)},J=0)$}} 
\; dy \, d\bm m \\
&= \E\left\{Y\indicator{A=a^{(0)}}w(a^{(0)},\ldots,a^{(t)},J=0) | L \right\}.
\end{align*}
Unbiased estimation using the inverse weighting procedure therefore depends on correctly specifying (i) a model for exposure conditional on covariates that converges in probability to its true value; and (ii) models for the (joint) distribution of the mediators, conditional on exposure and covariates, that are unbiased for the counterfactual distribution, i.e., $f(\bm M |A=a, L)$ converges in probability to its counterfactual distribution $f(\bm M_a |L)$, for $a=0,1$.

\section{Linear models}\label{sect:linear2M_POCeffects}

Estimation of the interventional indirect effects simplifies when the assumed models for the outcome and mediators are linear, because closed-form solutions for the estimators can be obtained by combining coefficients in the assumed models. For example, suppose that there are two mediators $M_1$ and $M_2$, and that the indirect effects are modified by a covariate $L$. For simplicity, we will assume in this example that there are no other confounders.
The (saturated) outcome mean model is:
\begin{align*}
&\E(Y|A=a, {M_{1}=m_1, M_{2}=m_2}, L=l) \\
&=\beta_0 + \beta_aa + \beta_1m_1 + \beta_2m_2 + \beta_ll \\
&\quad + \beta_{a,1}am_1 + \beta_{a,2}am_2  + \beta_{1,2}m_1m_2 + \beta_{a,1,2}am_1m_2 \\
&\quad + \beta_{a,l}al + \beta_{1,l}m_1l + \beta_{2,l}m_2l 
+ \beta_{1,2,l}m_1m_2l + \beta_{a,1,l}am_1l + \beta_{a,2,l}am_2l + \beta_{a,1,2,l}am_1m_2l.
\end{align*}
The (saturated) mean model for each mediator is:
\begin{align*}
\E(M_{s}|A=a, L=l) &= \alpha_{s,0} + \alpha_{s,a}a + \alpha_{s,l}l + \alpha_{s,a,l}al, \quad s=1,2.
\end{align*}
The (joint) indirect effect among individuals with covariate value $L=l$ is thus:
\begin{align}\label{eq:IE_2M_C}
{\rm IE_L} &= (\beta_1 + \beta_{1,l}l)(\alpha_{1,a}+\alpha_{1,a,l}l)
+ (\beta_2 + \beta_{2,l}l)(\alpha_{2,a}+\alpha_{2,a,l}l) 
+ (\beta_{1,2} + \beta_{1,2,l}l) \times
[\{\Sigma_{12}(1,l)-\Sigma_{12}(0,l)\} \notag\\
&\quad\quad+\{
(\alpha_{2,0} + \alpha_{2,l}l)(\alpha_{1a} + \alpha_{1,a,l}l)+
(\alpha_{1,0} + \alpha_{1,l}l)(\alpha_{2a} + \alpha_{2,a,l}l)+
(\alpha_{1,a} + \alpha_{1,a,l}l)(\alpha_{2a} + \alpha_{2,a,l}l)\}],
\end{align}
where the (conditional) covariance between the mediators is denoted by $\Sigma_{12}(a,l)=\cov(M_1,M_2|A=a, L=l)$. 
The indirect effect \eqref{eq:IE_2M_C} can then be decomposed into an indirect effect via $M_1$ and an indirect effect via $M_2$ as follows:
\begin{align*}
{\rm IE_L}_1 &= (\beta_1 + \beta_{1,l}l)(\alpha_{1,a}+\alpha_{1,a,l}l) 
+ (\beta_{1,2} + \beta_{1,2,l}l)
(\alpha_{2,0} + \alpha_{2,l}l)(\alpha_{1,a} + \alpha_{1,a,l}l), \\
{\rm IE_L}_2 &= (\beta_2 + \beta_{2,l}l)(\alpha_{2,a}+\alpha_{2,a,l}l)
+ (\beta_{1,2} + \beta_{1,2,l}l)
(\alpha_{1,0} + \alpha_{1,l}l + \alpha_{1,a} + \alpha_{1,a,l}l)(\alpha_{2,a} + \alpha_{2,a,l}l).
\end{align*}
The indirect effect via the mediators' mutual dependence, as defined in \eqref{eq:IEt_mutual_define}, is $(\beta_{1,2} + \beta_{1,2,l}l)\{\Sigma_{12}(1,l)-\Sigma_{12}(0,l)\}$.

\section{Simulation study details \label{sect:simstudies_details}}

\subsection{Study 1}

The observed data was generated as follows:
\begin{align*}
L &\sim {\cal N}(0,1) \\
A & \sim {\rm Bernoulli}\{\expit(0.7L)\}\\
M_1 &= A - 2L + \eps_1, \eps_1 \sim {\cal N}(0,1)\\
M_2 &= a_2A + e_{21}M_1 + L + \eps_2, \eps_2 \sim {\cal N}(0,1)\\
Y^\ast &= b_1 M_1 + M_2 + L\\
Y &\sim {\rm Bernoulli}\{\expit(Y^\ast)\}.
\end{align*}
The indirect effects via $M_1$ and via $M_2$ were non-zero (on a risk difference scale) when $b_1$ and $(a_2+e_{21})$ were non-zero respectively; the indirect effect via their mutual dependence, and the direct effect, were both zero.
We simulated 1000 datasets of sample size $n \in \{50, 250, 500\}$, with true values $b_1 \in \{0.0, 0.8\}$ and $(a_2,e_{21}) \in \{(0.0, 0.0), (0.0, 0.4), (0.0, 0.8), (0.4, 0.0), (0.8, 0.0)\}$. In particular, the indirect effect via $M_2$ was transmitted only via either a direct effect of $A$ on $M_2$ (so that the mediators were independent conditionally on $A$ and $L$), or an indirect effect of $A$ via $M_1$ (so that the mediators were correlated given $A$ and $L$).
The values of 0.4 and 0.8 were chosen merely to represent moderate and strong causal effects, without being so large as to induce perfect separation when fitting a logistic regression model for the outcome. 
Furthermore, the value of 0.8 ensured that the effect of $L$ on $M_2$ would be stronger than the effect of either $A$ or $M_1$ on $M_2$.

To demonstrate the invariance of the assumed causal ordering in yielding unbiased estimators, we (incorrectly) assumed $M_2$ to affect $M_1$ and merely factorized the joint mediator distribution as $f^a(\bm M|L) = f^a(M_2|L)f^a(M_1|M_2,L)$. 
For each simulated dataset, we assumed the mediators to be normally distributed as 
$M_2 | A=a, L \sim {\cal N}\{\E(M_2 | A=a, L), \sigma_2^2\}$ and $M_1 | A=a, M_2, L \sim {\cal N}\{\E(M_1 | A=a, M_2, L), \sigma_1^2\}$.
The following linear mediator models and logistic outcome model were fitted within each exposure group $A=a$ to estimate the (counterfactual) mediator densities or to impute the (potential) outcomes:
\begin{align*}
\E(M_2 | A=a, L) &= \alpha_{20}^a + \alpha_{2l}^a L, \\
\E(M_1 | A=a, M_2, L) &= \eta_{10}^a + \eta_{12}^a M_2 + \eta_{1l}^a L, \\
\logit\{\Pr(Y=1|A=a,M_1,M_2,L)\} &= \beta_0^a + \beta_1^a M_1 + \beta_2^a M_2 + \beta_l^a L.
\end{align*}
A propensity score model with a main effect for $L$ was fitted.
We fitted the effect model in \eqref{eq:effects_model_a00} (without any interactions terms with $L$, and with the logit link) using either the inverse weighting or Monte Carlo procedure.
The parameter estimates that encode the interventional (in)direct effects are displayed in Table~\ref{table:sim1_results}.
We discuss each of the two procedures in turn. 
The direct effect (``DE'') and joint indirect effect (``IE'') estimates were empirically unbiased under all the considered settings using the inverse weighting procedure, even when the factorization of the mediators' joint distribution was based on an incorrectly assumed causal ordering of the mediators. However, the estimates displayed biases at smaller sample sizes ($n=50$), which reduced to zero as the sample size increased.
The inverse weighting estimators of the separate indirect effects via each mediator $M_1$ (``IE1''), and $M_2$ (``IE2''), were unbiased when there was either a weak or no causal effect between the mediators ($e_{21}=0.4$ or $0$), but were biased when the effect was stronger ($e_{21}=0.8$). The biases, which persisted even in larger sample sizes, were due to unstable weights. The standard deviation of the weights (across all simulated datasets) for each row of the duplicated data in Table~\ref{table:duplicatedall_inverseweighting} are shown in Table~\ref{table:sim1_results_weights}. The weights in rows 1 to 3 that were used to calculate the indirect effects via each mediator were more variable when the effects among the mediators were stronger (e.g., $e_{21}=0.8$).
In contrast, the estimators of the interventional (in)direct effects using the Monte Carlo procedure were empirically unbiased under all the considered settings, although there were finite sample biases and larger empirical variability at smaller sample sizes that were reduced as the sample size increased.
Note that the Monte Carlo estimator as proposed in this paper requires no assumptions about the causal ordering of the mediators, because models for only the (marginal) mediator distribution, and not their joint distribution, are required.
Under the (true) linear and additive mediator models in this study, the Monte Carlo estimators would therefore be identical regardless of how the mediators' joint distribution was factorized.

\begingroup
\renewcommand\arraystretch{0.7}
\begin{longtable}{lrrrr|rrrrrr}
\caption{Average estimates (``est.'') and empirical standard errors (``ese'') of the interventional indirect and direct effects in simulation study 1. 
The estimators were obtained using either the inverse weighting (``IW'') procedure, or the imputation-based Monte Carlo (``MC'') procedure. 
The true values of the indirect effects were calculated by applying the Monte Carlo estimation method with the correct data-generating models to a population of $50,000$ individuals.
All results were rounded to two decimal places.}\label{table:sim1_results}\\
   & & & & & \multicolumn{2}{c}{IW} & \multicolumn{2}{c}{MC} \\  
Effect & $b_1$ & $e_{21}$ & $a_2$ & $n$ & true & est. & ese & est. & ese \\ 
  \hline\hline\endfirsthead  
\multicolumn{11}{c}{{\bfseries \tablename\ \thetable{} -- continued from previous page}} \\
   & & & & & \multicolumn{2}{c}{IW} & \multicolumn{2}{c}{MC} \\  
Effect & $b_1$ & $e_{21}$ & $a_2$ & $n$ & true & est. & ese & est. & ese \\ 
  \hline\hline\endhead
\hline \multicolumn{11}{c}{{Continued on next page}} \\ \hline
\endfoot
\hline \hline
\endlastfoot
  IE1 & 0.00 & 0.00 & 0.00 & 50 & 0.01 & -0.02 & 0.60 & -0.01 & 0.61 \\ 
  IE1 & 0.00 & 0.00 & 0.00 & 250 & 0.01 & 0.01 & 0.29 & 0.00 & 0.21 \\ 
  IE1 & 0.00 & 0.00 & 0.00 & 500 & 0.01 & -0.00 & 0.20 & -0.00 & 0.14 \\ 
  IE1 & 0.00 & 0.00 & 0.40 & 50 & 0.00 & 0.01 & 0.64 & 0.02 & 0.61 \\ 
  IE1 & 0.00 & 0.00 & 0.40 & 250 & 0.00 & 0.01 & 0.29 & 0.01 & 0.22 \\ 
  IE1 & 0.00 & 0.00 & 0.40 & 500 & 0.00 & -0.01 & 0.21 & 0.00 & 0.15 \\ 
  IE1 & 0.00 & 0.00 & 0.80 & 50 & 0.00 & 0.03 & 0.65 & 0.02 & 0.63 \\ 
  IE1 & 0.00 & 0.00 & 0.80 & 250 & 0.00 & -0.00 & 0.29 & 0.01 & 0.21 \\ 
  IE1 & 0.00 & 0.00 & 0.80 & 500 & 0.00 & -0.00 & 0.21 & -0.01 & 0.15 \\ 
  IE1 & 0.00 & 0.40 & 0.00 & 50 & -0.00 & 0.09 & 0.58 & 0.03 & 0.56 \\ 
  IE1 & 0.00 & 0.40 & 0.00 & 250 & -0.00 & 0.05 & 0.31 & 0.01 & 0.21 \\ 
  IE1 & 0.00 & 0.40 & 0.00 & 500 & -0.00 & 0.02 & 0.25 & 0.00 & 0.13 \\ 
  IE1 & 0.00 & 0.80 & 0.00 & 50 & 0.01 & 0.23 & 0.52 & 0.02 & 0.51 \\ 
  IE1 & 0.00 & 0.80 & 0.00 & 250 & 0.01 & 0.13 & 0.33 & 0.02 & 0.21 \\ 
  IE1 & 0.00 & 0.80 & 0.00 & 500 & 0.01 & 0.11 & 0.30 & 0.00 & 0.14 \\ 
  IE1 & 0.80 & 0.00 & 0.00 & 50 & 0.63 & 0.53 & 0.52 & 0.63 & 0.46 \\ 
  IE1 & 0.80 & 0.00 & 0.00 & 250 & 0.63 & 0.61 & 0.23 & 0.63 & 0.18 \\ 
  IE1 & 0.80 & 0.00 & 0.00 & 500 & 0.63 & 0.60 & 0.16 & 0.61 & 0.12 \\ 
  IE1 & 0.80 & 0.00 & 0.40 & 50 & 0.61 & 0.55 & 0.55 & 0.63 & 0.50 \\ 
  IE1 & 0.80 & 0.00 & 0.40 & 250 & 0.61 & 0.62 & 0.24 & 0.63 & 0.17 \\ 
  IE1 & 0.80 & 0.00 & 0.40 & 500 & 0.61 & 0.61 & 0.16 & 0.62 & 0.12 \\ 
  IE1 & 0.80 & 0.00 & 0.80 & 50 & 0.60 & 0.57 & 0.53 & 0.65 & 0.47 \\ 
  IE1 & 0.80 & 0.00 & 0.80 & 250 & 0.60 & 0.60 & 0.24 & 0.61 & 0.18 \\ 
  IE1 & 0.80 & 0.00 & 0.80 & 500 & 0.60 & 0.61 & 0.17 & 0.62 & 0.12 \\ 
  IE1 & 0.80 & 0.40 & 0.00 & 50 & 0.59 & 0.59 & 0.55 & 0.62 & 0.52 \\ 
  IE1 & 0.80 & 0.40 & 0.00 & 250 & 0.59 & 0.61 & 0.31 & 0.61 & 0.20 \\ 
  IE1 & 0.80 & 0.40 & 0.00 & 500 & 0.59 & 0.60 & 0.23 & 0.60 & 0.14 \\ 
  IE1 & 0.80 & 0.80 & 0.00 & 50 & 0.60 & 0.66 & 0.58 & 0.60 & 0.61 \\ 
  IE1 & 0.80 & 0.80 & 0.00 & 250 & 0.60 & 0.60 & 0.37 & 0.57 & 0.24 \\ 
  IE1 & 0.80 & 0.80 & 0.00 & 500 & 0.60 & 0.61 & 0.34 & 0.57 & 0.17 \\ 
  \hline  
  IE2 & 0.00 & 0.00 & 0.00 & 50 & 0.01 & 0.00 & 0.37 & 0.01 & 0.31 \\ 
  IE2 & 0.00 & 0.00 & 0.00 & 250 & 0.01 & -0.00 & 0.13 & -0.00 & 0.12 \\ 
  IE2 & 0.00 & 0.00 & 0.00 & 500 & 0.01 & -0.00 & 0.09 & -0.00 & 0.08 \\ 
  IE2 & 0.00 & 0.00 & 0.40 & 50 & 0.35 & 0.32 & 0.46 & 0.36 & 0.40 \\ 
  IE2 & 0.00 & 0.00 & 0.40 & 250 & 0.35 & 0.33 & 0.19 & 0.34 & 0.14 \\ 
  IE2 & 0.00 & 0.00 & 0.40 & 500 & 0.35 & 0.34 & 0.13 & 0.35 & 0.10 \\ 
  IE2 & 0.00 & 0.00 & 0.80 & 50 & 0.68 & 0.64 & 0.61 & 0.73 & 0.54 \\ 
  IE2 & 0.00 & 0.00 & 0.80 & 250 & 0.68 & 0.68 & 0.32 & 0.69 & 0.19 \\ 
  IE2 & 0.00 & 0.00 & 0.80 & 500 & 0.68 & 0.68 & 0.22 & 0.69 & 0.13 \\ 
  IE2 & 0.00 & 0.40 & 0.00 & 50 & 0.33 & 0.32 & 0.45 & 0.35 & 0.38 \\ 
  IE2 & 0.00 & 0.40 & 0.00 & 250 & 0.33 & 0.31 & 0.18 & 0.33 & 0.14 \\ 
  IE2 & 0.00 & 0.40 & 0.00 & 500 & 0.33 & 0.32 & 0.13 & 0.33 & 0.09 \\ 
  IE2 & 0.00 & 0.80 & 0.00 & 50 & 0.62 & 0.50 & 0.48 & 0.63 & 0.43 \\ 
  IE2 & 0.00 & 0.80 & 0.00 & 250 & 0.62 & 0.55 & 0.24 & 0.61 & 0.17 \\ 
  IE2 & 0.00 & 0.80 & 0.00 & 500 & 0.62 & 0.55 & 0.20 & 0.61 & 0.11 \\ 
  IE2 & 0.80 & 0.00 & 0.00 & 50 & -0.00 & 0.00 & 0.34 & -0.00 & 0.30 \\ 
  IE2 & 0.80 & 0.00 & 0.00 & 250 & -0.00 & 0.00 & 0.12 & 0.00 & 0.11 \\ 
  IE2 & 0.80 & 0.00 & 0.00 & 500 & -0.00 & -0.00 & 0.08 & -0.00 & 0.07 \\ 
  IE2 & 0.80 & 0.00 & 0.40 & 50 & 0.32 & 0.30 & 0.42 & 0.34 & 0.33 \\ 
  IE2 & 0.80 & 0.00 & 0.40 & 250 & 0.32 & 0.31 & 0.17 & 0.31 & 0.13 \\ 
  IE2 & 0.80 & 0.00 & 0.40 & 500 & 0.32 & 0.30 & 0.11 & 0.31 & 0.09 \\ 
  IE2 & 0.80 & 0.00 & 0.80 & 50 & 0.63 & 0.59 & 0.57 & 0.66 & 0.44 \\ 
  IE2 & 0.80 & 0.00 & 0.80 & 250 & 0.63 & 0.61 & 0.24 & 0.62 & 0.16 \\ 
  IE2 & 0.80 & 0.00 & 0.80 & 500 & 0.63 & 0.63 & 0.18 & 0.63 & 0.12 \\ 
  IE2 & 0.80 & 0.40 & 0.00 & 50 & 0.31 & 0.28 & 0.41 & 0.31 & 0.35 \\ 
  IE2 & 0.80 & 0.40 & 0.00 & 250 & 0.31 & 0.31 & 0.18 & 0.31 & 0.13 \\ 
  IE2 & 0.80 & 0.40 & 0.00 & 500 & 0.31 & 0.30 & 0.13 & 0.30 & 0.09 \\ 
  IE2 & 0.80 & 0.80 & 0.00 & 50 & 0.57 & 0.52 & 0.52 & 0.58 & 0.47 \\ 
  IE2 & 0.80 & 0.80 & 0.00 & 250 & 0.57 & 0.56 & 0.27 & 0.58 & 0.18 \\ 
  IE2 & 0.80 & 0.80 & 0.00 & 500 & 0.57 & 0.55 & 0.23 & 0.58 & 0.13 \\ 
  \hline  
  Joint IE & 0.00 & 0.00 & 0.00 & 50 & 0.03 & -0.02 & 0.53 & -0.01 & 0.58 \\ 
  Joint IE & 0.00 & 0.00 & 0.00 & 250 & 0.03 & 0.01 & 0.24 & -0.00 & 0.21 \\ 
  Joint IE & 0.00 & 0.00 & 0.00 & 500 & 0.03 & -0.01 & 0.17 & -0.00 & 0.14 \\ 
  Joint IE & 0.00 & 0.00 & 0.40 & 50 & 0.34 & 0.32 & 0.62 & 0.35 & 0.62 \\ 
  Joint IE & 0.00 & 0.00 & 0.40 & 250 & 0.34 & 0.34 & 0.25 & 0.35 & 0.20 \\ 
  Joint IE & 0.00 & 0.00 & 0.40 & 500 & 0.34 & 0.34 & 0.18 & 0.35 & 0.14 \\ 
  Joint IE & 0.00 & 0.00 & 0.80 & 50 & 0.68 & 0.64 & 0.65 & 0.73 & 0.70 \\ 
  Joint IE & 0.00 & 0.00 & 0.80 & 250 & 0.68 & 0.66 & 0.32 & 0.69 & 0.22 \\ 
  Joint IE & 0.00 & 0.00 & 0.80 & 500 & 0.68 & 0.67 & 0.23 & 0.68 & 0.16 \\ 
  Joint IE & 0.00 & 0.40 & 0.00 & 50 & 0.33 & 0.32 & 0.53 & 0.36 & 0.52 \\ 
  Joint IE & 0.00 & 0.40 & 0.00 & 250 & 0.33 & 0.31 & 0.23 & 0.33 & 0.18 \\ 
  Joint IE & 0.00 & 0.40 & 0.00 & 500 & 0.33 & 0.32 & 0.16 & 0.33 & 0.13 \\ 
  Joint IE & 0.00 & 0.80 & 0.00 & 50 & 0.63 & 0.56 & 0.47 & 0.65 & 0.47 \\ 
  Joint IE & 0.00 & 0.80 & 0.00 & 250 & 0.63 & 0.60 & 0.21 & 0.62 & 0.17 \\ 
  Joint IE & 0.00 & 0.80 & 0.00 & 500 & 0.63 & 0.61 & 0.15 & 0.61 & 0.13 \\ 
  Joint IE & 0.80 & 0.00 & 0.00 & 50 & 0.63 & 0.50 & 0.49 & 0.62 & 0.49 \\ 
  Joint IE & 0.80 & 0.00 & 0.00 & 250 & 0.63 & 0.60 & 0.22 & 0.63 & 0.18 \\ 
  Joint IE & 0.80 & 0.00 & 0.00 & 500 & 0.63 & 0.60 & 0.15 & 0.61 & 0.12 \\ 
  Joint IE & 0.80 & 0.00 & 0.40 & 50 & 0.93 & 0.84 & 0.52 & 0.99 & 0.50 \\ 
  Joint IE & 0.80 & 0.00 & 0.40 & 250 & 0.93 & 0.91 & 0.22 & 0.94 & 0.18 \\ 
  Joint IE & 0.80 & 0.00 & 0.40 & 500 & 0.93 & 0.90 & 0.15 & 0.92 & 0.13 \\ 
  Joint IE & 0.80 & 0.00 & 0.80 & 50 & 1.24 & 1.10 & 0.57 & 1.32 & 0.55 \\ 
  Joint IE & 0.80 & 0.00 & 0.80 & 250 & 1.24 & 1.20 & 0.26 & 1.24 & 0.20 \\ 
  Joint IE & 0.80 & 0.00 & 0.80 & 500 & 1.24 & 1.22 & 0.19 & 1.25 & 0.13 \\ 
  Joint IE & 0.80 & 0.40 & 0.00 & 50 & 0.86 & 0.74 & 0.50 & 0.88 & 0.54 \\ 
  Joint IE & 0.80 & 0.40 & 0.00 & 250 & 0.86 & 0.85 & 0.23 & 0.87 & 0.20 \\ 
  Joint IE & 0.80 & 0.40 & 0.00 & 500 & 0.86 & 0.83 & 0.16 & 0.85 & 0.13 \\ 
  Joint IE & 0.80 & 0.80 & 0.00 & 50 & 1.09 & 0.95 & 0.54 & 1.12 & 0.57 \\ 
  Joint IE & 0.80 & 0.80 & 0.00 & 250 & 1.09 & 1.03 & 0.23 & 1.09 & 0.22 \\ 
  Joint IE & 0.80 & 0.80 & 0.00 & 500 & 1.09 & 1.04 & 0.17 & 1.08 & 0.15 \\
  \hline
DE & 0.00 & 0.00 & 0.00 & 50 & 0.00 & -0.06 & 0.98 & -0.04 & 0.93 \\ 
  DE & 0.00 & 0.00 & 0.00 & 250 & 0.00 & -0.04 & 0.39 & -0.02 & 0.34 \\ 
  DE & 0.00 & 0.00 & 0.00 & 500 & 0.00 & -0.00 & 0.27 & -0.01 & 0.23 \\ 
  DE & 0.00 & 0.00 & 0.40 & 50 & -0.02 & 0.06 & 1.06 & 0.02 & 0.97 \\ 
  DE & 0.00 & 0.00 & 0.40 & 250 & -0.02 & 0.01 & 0.39 & -0.00 & 0.34 \\ 
  DE & 0.00 & 0.00 & 0.40 & 500 & -0.02 & -0.01 & 0.25 & -0.02 & 0.23 \\ 
  DE & 0.00 & 0.00 & 0.80 & 50 & -0.02 & 0.18 & 1.41 & 0.08 & 1.42 \\ 
  DE & 0.00 & 0.00 & 0.80 & 250 & -0.02 & 0.05 & 0.42 & 0.01 & 0.34 \\ 
  DE & 0.00 & 0.00 & 0.80 & 500 & -0.02 & 0.03 & 0.30 & 0.01 & 0.24 \\ 
  DE & 0.00 & 0.40 & 0.00 & 50 & 0.00 & 0.05 & 0.90 & -0.00 & 0.81 \\ 
  DE & 0.00 & 0.40 & 0.00 & 250 & 0.00 & 0.01 & 0.33 & -0.00 & 0.29 \\ 
  DE & 0.00 & 0.40 & 0.00 & 500 & 0.00 & -0.00 & 0.23 & -0.01 & 0.21 \\ 
  DE & 0.00 & 0.80 & 0.00 & 50 & -0.03 & 0.12 & 0.75 & 0.02 & 0.65 \\ 
  DE & 0.00 & 0.80 & 0.00 & 250 & -0.03 & 0.04 & 0.30 & 0.02 & 0.26 \\ 
  DE & 0.00 & 0.80 & 0.00 & 500 & -0.03 & 0.01 & 0.21 & 0.00 & 0.19 \\ 
  DE & 0.80 & 0.00 & 0.00 & 50 & -0.04 & 0.16 & 0.81 & 0.03 & 0.72 \\ 
  DE & 0.80 & 0.00 & 0.00 & 250 & -0.04 & 0.02 & 0.30 & -0.00 & 0.26 \\ 
  DE & 0.80 & 0.00 & 0.00 & 500 & -0.04 & 0.01 & 0.20 & 0.00 & 0.18 \\ 
  DE & 0.80 & 0.00 & 0.40 & 50 & 0.01 & 0.20 & 0.83 & 0.02 & 0.73 \\ 
  DE & 0.80 & 0.00 & 0.40 & 250 & 0.01 & 0.02 & 0.32 & -0.01 & 0.28 \\ 
  DE & 0.80 & 0.00 & 0.40 & 500 & 0.01 & 0.03 & 0.21 & 0.01 & 0.19 \\ 
  DE & 0.80 & 0.00 & 0.80 & 50 & -0.01 & 0.42 & 1.45 & 0.14 & 1.50 \\ 
  DE & 0.80 & 0.00 & 0.80 & 250 & -0.01 & 0.06 & 0.34 & 0.00 & 0.29 \\ 
  DE & 0.80 & 0.00 & 0.80 & 500 & -0.01 & 0.03 & 0.26 & -0.00 & 0.21 \\ 
  DE & 0.80 & 0.40 & 0.00 & 50 & -0.01 & 0.25 & 0.82 & 0.07 & 0.72 \\ 
  DE & 0.80 & 0.40 & 0.00 & 250 & -0.01 & 0.02 & 0.30 & -0.01 & 0.26 \\ 
  DE & 0.80 & 0.40 & 0.00 & 500 & -0.01 & 0.02 & 0.21 & -0.00 & 0.18 \\ 
  DE & 0.80 & 0.80 & 0.00 & 50 & -0.01 & 0.29 & 0.82 & 0.04 & 0.69 \\ 
  DE & 0.80 & 0.80 & 0.00 & 250 & -0.01 & 0.03 & 0.33 & -0.01 & 0.28 \\ 
  DE & 0.80 & 0.80 & 0.00 & 500 & -0.01 & 0.03 & 0.23 & 0.01 & 0.18 \\   
   \hline  
\end{longtable}
\endgroup

\begin{table}[ht]
\renewcommand{\arraystretch}{1.2}
\centering
\caption{Standard deviation of the weights in each row of the duplicated data for the inverse weighting estimator in study 1. 
All results were rounded to two decimal places. \label{table:sim1_results_weights}}
\begin{tabular}{rrrr|rrrrr}
  \hline
   & & & & \multicolumn{5}{c}{Row in Table~\ref{table:duplicatedall_inverseweighting}}  \\
$b_1$ & $e_{21}$ & $a_2$ & $n$ & 1 & 2 & 3 & 4 & 5 \\ 
  \hline
0.00 & 0.00 & 0.00 & 50 & 1.10 & 2.73 & 2.77 & 2.73 & 1.01 \\ 
  0.00 & 0.00 & 0.00 & 250 & 0.83 & 3.10 & 3.28 & 3.27 & 0.80 \\ 
  0.00 & 0.00 & 0.00 & 500 & 0.79 & 3.18 & 3.23 & 3.18 & 0.78 \\ 
  0.00 & 0.00 & 0.40 & 50 & 1.15 & 2.84 & 3.11 & 3.19 & 1.04 \\ 
  0.00 & 0.00 & 0.40 & 250 & 0.82 & 3.08 & 3.62 & 3.58 & 0.79 \\ 
  0.00 & 0.00 & 0.40 & 500 & 0.78 & 3.17 & 3.42 & 3.49 & 0.76 \\ 
  0.00 & 0.00 & 0.80 & 50 & 1.13 & 2.51 & 3.40 & 3.36 & 1.04 \\ 
  0.00 & 0.00 & 0.80 & 250 & 0.83 & 3.15 & 5.05 & 5.35 & 0.80 \\ 
  0.00 & 0.00 & 0.80 & 500 & 0.79 & 3.07 & 4.48 & 4.70 & 0.77 \\ 
  0.00 & 0.40 & 0.00 & 50 & 1.21 & 2.86 & 3.19 & 3.06 & 1.06 \\ 
  0.00 & 0.40 & 0.00 & 250 & 1.36 & 4.61 & 3.64 & 3.58 & 0.80 \\ 
  0.00 & 0.40 & 0.00 & 500 & 1.35 & 5.59 & 4.11 & 3.12 & 0.78 \\ 
  0.00 & 0.80 & 0.00 & 50 & 1.67 & 2.99 & 2.82 & 2.99 & 0.96 \\ 
  0.00 & 0.80 & 0.00 & 250 & 3.59 & 10.89 & 5.52 & 3.36 & 0.81 \\ 
  0.00 & 0.80 & 0.00 & 500 & 17.53 & 13.79 & 12.14 & 3.15 & 0.77 \\ 
  0.80 & 0.00 & 0.00 & 50 & 1.14 & 2.86 & 3.02 & 2.92 & 1.00 \\ 
  0.80 & 0.00 & 0.00 & 250 & 0.83 & 3.22 & 3.19 & 3.25 & 0.80 \\ 
  0.80 & 0.00 & 0.00 & 500 & 0.79 & 3.07 & 3.16 & 3.10 & 0.78 \\ 
  0.80 & 0.00 & 0.40 & 50 & 1.12 & 2.69 & 2.99 & 3.19 & 1.06 \\ 
  0.80 & 0.00 & 0.40 & 250 & 0.82 & 3.18 & 3.96 & 3.60 & 0.80 \\ 
  0.80 & 0.00 & 0.40 & 500 & 0.78 & 3.01 & 3.55 & 3.77 & 0.77 \\ 
  0.80 & 0.00 & 0.80 & 50 & 1.06 & 2.89 & 3.99 & 3.61 & 0.96 \\ 
  0.80 & 0.00 & 0.80 & 250 & 0.83 & 3.09 & 4.58 & 4.58 & 0.81 \\ 
  0.80 & 0.00 & 0.80 & 500 & 0.80 & 3.12 & 5.16 & 4.93 & 0.78 \\ 
  0.80 & 0.40 & 0.00 & 50 & 1.28 & 2.85 & 2.75 & 2.90 & 1.00 \\ 
  0.80 & 0.40 & 0.00 & 250 & 1.43 & 4.87 & 3.72 & 3.30 & 0.81 \\ 
  0.80 & 0.40 & 0.00 & 500 & 1.39 & 5.12 & 3.98 & 3.24 & 0.76 \\ 
  0.80 & 0.80 & 0.00 & 50 & 1.77 & 3.19 & 2.82 & 2.70 & 0.97 \\ 
  0.80 & 0.80 & 0.00 & 250 & 3.12 & 7.10 & 4.37 & 3.09 & 0.80 \\ 
  0.80 & 0.80 & 0.00 & 500 & 6.86 & 12.82 & 6.53 & 3.08 & 0.77 \\ 
   \hline
\end{tabular}
\end{table}

\clearpage

\subsection{Study 2}

The observed data was generated as follows:
\begin{align*}
L_1 &\sim {\cal N}(0,1) \\
L_2 &\sim {\rm Bernoulli}(0.1)\\
A & \sim {\rm Bernoulli}\{\expit(0.7 L_2)\}\\
M_1 &= AL_2 - L_2 + L_1 + \eps_1, \eps_1 \sim {\cal N}(0,1)\\
M_2 &= a_2A + e_{21}M_1 + L_1 + L_2 + \eps_2, \eps_2 \sim {\cal N}(0,1)\\
Y &= b_1 M_1 + M_2 + L_1 + L_2 + \eps_y, \eps_y \sim {\cal N}(0,1).
\end{align*}
This study differed from study 1 in that (i) there were two confounders $L_1$ and $L_2$, (ii) the effect of exposure $A$ on $M_1$ was restricted to individuals with $L_2=1$, and (iii) the outcome was continuous.
The indirect effect via $M_1$, which equalled $b_1$ among individuals with $L_2=1$ but was zero otherwise (regardless of the assumed value of $b_1$), was thus modified by $L_2$.
Furthermore, when $e_{21}$ was non-zero, the indirect effect via $M_2$ was modified by $L_2$ through the dependence of $M_2$ on $M_1$; e.g., the indirect effect via $M_2$ equalled $a_2$ among individuals with $L_2=0$, or $a_2+e_{21}$ among individuals with $L_2=1$.
We assumed continuous mediators and outcome so that closed form expressions of the interventional indirect effects could be determined in terms of the data-generating model coefficients as described in Appendix~\ref{sect:linear2M_POCeffects}.

We simulated 1000 datasets of sample size $n=100$, with true values $b_1 \in \{0.0, 1.6\}$ and  $(a_2,e_{21}) \in \{(0.0, 0.0), (0.0, 1.6), (1.6, 0.0)\}$.
The value of 1.6 was chosen so that the exposure effects on $M_2$, and on $Y$, would be stronger than the effects of either confounder on $M_2$, and on $Y$.
For each simulated dataset, we assumed the mediators to be normally distributed as 
$M_1 | A=a, L_1, L_2 \sim {\cal N}\{\E(M_2 | A=a, L_1, L_2), \sigma_1^2\}$ and $M_2 | A=a, M_1, L_1, L_2 \sim {\cal N}\{\E(M_2 | A=a, M_1, L_1, L_2), \sigma_2^2\}$.
The following (correctly-specified) linear mediator and outcome models were fitted within each exposure group $A=a$ to estimate the (counterfactual) mediator densities or to impute the (potential) outcomes:
\begin{align*}
\E(M_1 | A=a, L_1, L_2) &= \alpha_{10}^a + \alpha_{1,l1}^a L_1 + \alpha_{1,l2}^a L_2, \\
\E(M_2 | A=a, M_1, L_1, L_2) &= \eta_{20}^a + \eta_{21}^a M_1 + \eta_{21,l2}^a M_1L_2 + \eta_{2,l1}^a L_1 + \eta_{2,l2}^a L_2, \\
\E(Y|A=a,M_1,M_2,L_1, L_2) &= \beta_0^a + \beta_1^a M_1 + \beta_2^a M_2 
+ \beta_{1,l2}^a M_1L_2 + \beta_{2,l2}^a M_2L_2 + \beta_{l1}^a L_1 + \beta_{l2}^a L_2.
\end{align*}
The mediator-covariate interaction terms were included to allow the effects of the mediators on the outcome to differ depending on $L_2$.
A (correctly-specified) propensity score model with main effects for $L_1$ and $L_2$ was fitted.
We fitted the effect model in \eqref{eq:effects_model_a00} (with the identity link) using either the inverse weighting or Monte Carlo procedure.
The resulting parameter estimates that indexed the effect model are displayed in Table~\ref{table:sim2_results}.
Furthermore, we assessed the empirical coverage of nonparametric percentile bootstrap 95\% confidence intervals, constructed by randomly resampling observations with replacement and repeating the estimation procedures for each bootstrap sample.
To reduce the computing time needed to carry out the bootstrap procedure for each generated dataset, we only considered settings with a sample size of $n=100$, and limited the number of bootstrap resamples to just 100. In practice, a larger number of bootstrap resamples should be used to better represent the (tails of the) bootstrap distribution.

The results were similar to those in simulation study 1.
Estimators using the inverse weighting procedure were empirically unbiased only when there were no exposure effects on the mediators, which ensured that the weights were most stable.
In all other settings, the inverse weighting estimators were biased, and the coverage of the confidence intervals were below their nominal levels.
In contrast, estimators using the Monte Carlo procedure were empirically unbiased, and the confidence intervals approximately attained the nominal coverage, under all considered settings.

\begingroup
\renewcommand\arraystretch{0.7}
\begin{longtable}{rrrlr|rrrrrr}
\caption{Average values (``est.''), empirical standard errors (``ese''), and coverage of the 95\% confidence intervals (``CP'') for the interventional indirect effect estimators in simulation study 2. 
The estimators were obtained using either the inverse weighting (``IW'') procedure, or the imputation-based Monte Carlo (``MC'') procedure.
Indirect effects among individuals with $L_2=1$ ($L_2=0$) are with(out) ``\_L2'' appended in the label.
All results were rounded to two decimal places.}\label{table:sim2_results}\\
  \hline
   & & & & & \multicolumn{2}{c}{IW} & \multicolumn{2}{c}{MC} & IW & MC \\
$b_1$ & $e_{21}$ & $a_2$ & Effect & true & est. & ese & est. & ese & \multicolumn{2}{c}{CP} \\ 
  \hline\hline\endfirsthead
\multicolumn{11}{c}{{\bfseries \tablename\ \thetable{} -- continued from previous page}} \\
   & & & & & \multicolumn{2}{c}{IW} & \multicolumn{2}{c}{MC} & IW & MC \\
$b_1$ & $e_{21}$ & $a_2$ & Effect & true & est. & ese & est. & ese & \multicolumn{2}{c}{CP} \\ 
  \hline\hline\endhead
\hline \multicolumn{11}{c}{{Continued on next page}} \\ \hline
\endfoot
\hline \hline
\endlastfoot
0.00 & 0.00 & 0.00 & IE1 & 0.00 & -0.00 & 0.11 & -0.00 & 0.11 & 1.00 & 1.00 \\ 
  0.00 & 0.00 & 1.60 & IE1 & 0.00 & -0.00 & 0.10 & 0.00 & 0.11 & 1.00 & 1.00 \\ 
  0.00 & 1.60 & 0.00 & IE1 & 0.00 & 0.00 & 0.22 & 0.00 & 0.22 & 0.97 & 1.00 \\ 
  1.60 & 0.00 & 0.00 & IE1 & 0.00 & -0.00 & 0.45 & -0.00 & 0.51 & 0.96 & 0.94 \\ 
  1.60 & 0.00 & 1.60 & IE1 & 0.00 & 0.00 & 0.45 & -0.00 & 0.51 & 0.96 & 0.94 \\ 
  1.60 & 1.60 & 0.00 & IE1 & 0.00 & 0.03 & 0.50 & 0.03 & 0.51 & 0.95 & 0.97 \\ 
  \hline
  0.00 & 0.00 & 0.00 & IE1\_L2 & 0.00 & 0.01 & 0.32 & 0.00 & 0.29 & 0.95 & 0.96 \\ 
  0.00 & 0.00 & 1.60 & IE1\_L2 & 0.00 & -0.02 & 0.32 & -0.01 & 0.30 & 0.96 & 0.96 \\ 
  0.00 & 1.60 & 0.00 & IE1\_L2 & 0.00 & 0.63 & 0.46 & -0.01 & 0.61 & 0.54 & 0.97 \\ 
  1.60 & 0.00 & 0.00 & IE1\_L2 & 1.60 & 1.24 & 0.67 & 1.59 & 0.59 & 0.81 & 0.95 \\ 
  1.60 & 0.00 & 1.60 & IE1\_L2 & 1.60 & 1.23 & 0.68 & 1.56 & 0.59 & 0.80 & 0.94 \\ 
  1.60 & 1.60 & 0.00 & IE1\_L2 & 1.60 & 1.54 & 0.86 & 1.63 & 0.82 & 0.93 & 0.95 \\ 
  \hline
  0.00 & 0.00 & 0.00 & IE2 & 0.00 & -0.01 & 0.27 & -0.02 & 0.31 & 0.97 & 0.94 \\ 
  0.00 & 0.00 & 1.60 & IE2 & 1.60 & 1.17 & 0.56 & 1.60 & 0.47 & 0.69 & 0.93 \\ 
  0.00 & 1.60 & 0.00 & IE2 & 0.00 & -0.01 & 0.37 & -0.01 & 0.59 & 0.95 & 0.94 \\ 
  1.60 & 0.00 & 0.00 & IE2 & 0.00 & 0.01 & 0.29 & 0.01 & 0.33 & 0.98 & 0.97 \\ 
  1.60 & 0.00 & 1.60 & IE2 & 1.60 & 1.17 & 0.70 & 1.59 & 0.47 & 0.78 & 0.95 \\ 
  1.60 & 1.60 & 0.00 & IE2 & 0.00 & 0.02 & 0.48 & 0.01 & 0.58 & 0.95 & 0.95 \\ 
  \hline
  0.00 & 0.00 & 0.00 & IE2\_L2 & 0.00 & -0.01 & 0.31 & -0.00 & 0.35 & 0.98 & 0.93 \\ 
  0.00 & 0.00 & 1.60 & IE2\_L2 & 1.60 & 1.03 & 0.66 & 1.60 & 0.54 & 0.61 & 0.93 \\ 
  0.00 & 1.60 & 0.00 & IE2\_L2 & 1.60 & 0.82 & 0.60 & 1.58 & 0.77 & 0.50 & 0.93 \\ 
  1.60 & 0.00 & 0.00 & IE2\_L2 & 0.00 & -0.02 & 0.33 & -0.01 & 0.36 & 0.99 & 0.97 \\ 
  1.60 & 0.00 & 1.60 & IE2\_L2 & 1.60 & 0.82 & 0.78 & 1.59 & 0.58 & 0.55 & 0.95 \\ 
  1.60 & 1.60 & 0.00 & IE2\_L2 & 1.60 & 1.22 & 0.80 & 1.60 & 0.76 & 0.81 & 0.95 \\ 
  \hline
  0.00 & 0.00 & 0.00 & Joint IE & 0.00 & -0.01 & 0.27 & -0.02 & 0.30 & 0.96 & 0.93 \\ 
  0.00 & 0.00 & 1.60 & Joint IE & 1.60 & 1.22 & 0.42 & 1.59 & 0.39 & 0.74 & 0.92 \\ 
  0.00 & 1.60 & 0.00 & Joint IE & 0.00 & -0.03 & 0.54 & -0.02 & 0.56 & 0.96 & 0.93 \\ 
  1.60 & 0.00 & 0.00 & Joint IE & 0.00 & 0.01 & 0.52 & 0.01 & 0.55 & 0.96 & 0.92 \\ 
  1.60 & 0.00 & 1.60 & Joint IE & 1.60 & 1.20 & 0.66 & 1.58 & 0.61 & 0.85 & 0.93 \\ 
  1.60 & 1.60 & 0.00 & Joint IE & 0.00 & 0.05 & 0.92 & 0.05 & 0.94 & 0.97 & 0.93 \\ 
  \hline
  0.00 & 0.00 & 0.00 & Joint IE\_L2 & 0.00 & 0.00 & 0.36 & -0.00 & 0.37 & 0.96 & 0.93 \\ 
  0.00 & 0.00 & 1.60 & Joint IE\_L2 & 1.60 & 1.17 & 0.46 & 1.59 & 0.42 & 0.74 & 0.94 \\ 
  0.00 & 1.60 & 0.00 & Joint IE\_L2 & 1.60 & 1.32 & 0.62 & 1.57 & 0.62 & 0.90 & 0.93 \\ 
  1.60 & 0.00 & 0.00 & Joint IE\_L2 & 1.60 & 1.35 & 0.60 & 1.60 & 0.62 & 0.92 & 0.93 \\ 
  1.60 & 0.00 & 1.60 & Joint IE\_L2 & 3.20 & 2.39 & 0.68 & 3.14 & 0.65 & 0.70 & 0.92 \\ 
  1.60 & 1.60 & 0.00 & Joint IE\_L2 & 3.20 & 2.71 & 1.06 & 3.21 & 1.05 & 0.91 & 0.93 \\ 
  \hline
  0.00 & 0.00 & 0.00 & DE & 0.00 & -0.01 & 0.26 & -0.00 & 0.24 & 0.95 & 0.94 \\ 
  0.00 & 0.00 & 1.60 & DE & 0.00 & 0.42 & 0.37 & 0.02 & 0.32 & 0.64 & 0.94 \\ 
  0.00 & 1.60 & 0.00 & DE & 0.00 & 0.13 & 0.32 & 0.00 & 0.25 & 0.93 & 0.93 \\ 
  1.60 & 0.00 & 0.00 & DE & 0.00 & 0.12 & 0.32 & -0.00 & 0.24 & 0.92 & 0.94 \\ 
  1.60 & 0.00 & 1.60 & DE & 0.00 & 0.57 & 0.48 & 0.01 & 0.31 & 0.63 & 0.94 \\ 
  1.60 & 1.60 & 0.00 & DE & 0.00 & 0.26 & 0.48 & 0.00 & 0.24 & 0.91 & 0.94 \\ 
   \hline\hline
\end{longtable}
\endgroup

\subsection{Study 3}

The observed data was generated as follows:
\begin{align*}
L_1 &\sim {\cal N}(0,1) \\
L_2 &\sim {\rm Bernoulli}(0.1)\\
A & \sim {\rm Bernoulli}\{\expit(0.7 L_2)\}\\
M_1 &= AL_2 - L_2 + L_1 + \eps_1, \eps_1 \sim {\cal N}(0,1)\\
M_2 &= a_2AM_1 + e_{21}M_1 + L_1 + L_2 + \eps_2, \eps_2 \sim {\cal N}(0,1)\\
Y^\ast &= b_1 M_1 + M_2 - 0.4M_1M_2 + L_1 + L_2\\
Y &\sim {\rm Bernoulli}\{\expit(Y^\ast)\}.
\end{align*}
This study combined elements of the data-generating processes from the previous studies, such as the exposure and mediator models from study 2, and the model for a binary outcome from study 1. In addition, the effect of $M_1$ on $M_2$ was now dependent on exposure $A$ due to the $A-M_1$ interaction term in the model for $M_2$, and there was a mediator-mediator interaction term in the outcome model.
We simulated 1000 datasets of sample size $n \in \{50, 250, 500\}$, with true values $b_1=0$ and $(a_2,e_{21}) \in \{(0.0, 0.0), (0.8, 0.0), (0.0, 0.8)\}$. These settings were a subset of those considered in study 1; in particular the setting of $b_1=a_2=0.0, e_{21}=0.8$ resulted in the highly variable weights for the inverse weighting estimator.
For each simulated dataset, the propensity score and correctly-specified mediator models (as described in the previous study) were fitted.
We fitted the effect model in \eqref{eq:effects_model_a00} (with the logit link) using either the inverse weighting or Monte Carlo procedure.
For the Monte Carlo procedure, the following outcome model was fitted:
\begin{align*}
\logit\{\Pr(Y=1|A=a,M_1,M_2,L_1, L_2)\} &= \beta_0^a + \beta_1^a M_1 + \beta_2^a M_2 
+ \beta_{1,l2}^a M_1L_2 + \beta_{2,l2}^a M_2L_2 \\
&\quad + \beta_{12}^a M_1M_2 + \beta_{12,l2}^a M_1M_2L_2 + \beta_{l1}^a L_1 + \beta_{l2}^a L_2.
\end{align*}
Furthermore, to assess the biases from incorrectly specifying the outcome model, we considering an additional Monte Carlo estimator that omitted all mediator-mediator interaction terms. Note that no outcome model is required for the proposed inverse weighting estimator.

The results of this simulation study are displayed in Table~\ref{table:sim3_results}.
We briefly describe the conclusions which were similar to those in the previous simulation studies.
Estimators using the inverse weighting procedure were empirically unbiased only when there were no exposure effects on the mediators and the weights were most stable.
In all other settings, the inverse weighting estimators were biased.
Estimators using the Monte Carlo procedure were empirically unbiased when the outcome model was correctly specified, under all considered settings.
But omitting the mediator-mediator interaction terms from the fitted outcome model resulted in biased estimates of the interventional indirect effects, especially in smaller sample sizes.

\begingroup
\renewcommand\arraystretch{0.7}
\begin{longtable}{lrrrr|rrrrrr}
\caption{Average estimates (``est.'') and empirical standard errors (``ese'') of the interventional indirect effects in simulation study 3. 
The estimators were obtained using either the inverse weighting (``IW'') procedure, or the imputation-based Monte Carlo (``MC'') procedure.
The assumed outcome ($Y$) model in the latter procedure was either incorrect (``$Y$ mis.'') or correct.
Indirect effects among individuals with $L_2=1$ ($L_2=0$) are with(out) ``\_L2'' appended in the label.
The true values of the indirect effects were calculating by applying the Monte Carlo estimation method with the correct data-generating models to a population of $5 \times 10^4$ individuals.
All results were rounded to two decimal places.}\label{table:sim3_results}\\
  \hline
   & & & & & \multicolumn{2}{c}{IW} & \multicolumn{2}{c}{MC ($Y$ mis.)} & \multicolumn{2}{c}{MC} \\
Effect & $e_{21}$ & $a_2$ & $n$ & true & est. & ese & est. & ese & est. & ese \\ 
  \hline\hline\endfirsthead
  \multicolumn{11}{c}{{\bfseries \tablename\ \thetable{} -- continued from previous page}} \\
   & & & & & \multicolumn{2}{c}{IW} & \multicolumn{2}{c}{MC ($Y$ mis.)} & \multicolumn{2}{c}{MC} \\
Effect & $e_{21}$ & $a_2$ & $n$ & true & est. & ese & est. & ese & est. & ese \\ 
  \hline\hline\endhead
\hline \multicolumn{11}{c}{{Continued on next page}} \\ \hline
\endfoot
\hline \hline
\endlastfoot
  IE1 & 0.00 & 0.00 & 50 & 0.00 & 0.02 & 0.39 & 0.04 & 0.39 & 0.05 & 0.37 \\ 
  IE1 & 0.00 & 0.00 & 250 & 0.00 & -0.00 & 0.09 & 0.00 & 0.06 & -0.00 & 0.07 \\ 
  IE1 & 0.00 & 0.00 & 500 & 0.00 & 0.00 & 0.05 & 0.01 & 0.03 & 0.00 & 0.04 \\ 
  IE1 & 0.00 & 0.80 & 50 & 0.00 & -0.01 & 0.43 & 0.02 & 0.43 & 0.02 & 0.40 \\ 
  IE1 & 0.00 & 0.80 & 250 & 0.00 & -0.01 & 0.10 & -0.00 & 0.07 & -0.00 & 0.07 \\ 
  IE1 & 0.00 & 0.80 & 500 & 0.00 & -0.01 & 0.07 & -0.00 & 0.04 & 0.00 & 0.04 \\ 
  IE1 & 0.80 & 0.00 & 50 & 0.02 & 0.03 & 0.47 & 0.06 & 0.46 & 0.03 & 0.42 \\ 
  IE1 & 0.80 & 0.00 & 250 & 0.02 & 0.03 & 0.19 & 0.03 & 0.09 & 0.03 & 0.10 \\ 
  IE1 & 0.80 & 0.00 & 500 & 0.02 & 0.02 & 0.13 & 0.02 & 0.05 & 0.02 & 0.06 \\ 
  \hline
  IE1\_L2 & 0.00 & 0.00 & 50 & -0.13 & 0.02 & 1.07 & -0.30 & 2.07 & -0.45 & 1.98 \\ 
  IE1\_L2 & 0.00 & 0.00 & 250 & -0.13 & -0.03 & 0.51 & -0.07 & 0.37 & -0.12 & 0.40 \\ 
  IE1\_L2 & 0.00 & 0.00 & 500 & -0.13 & -0.07 & 0.35 & -0.07 & 0.23 & -0.10 & 0.25 \\ 
  IE1\_L2 & 0.00 & 0.80 & 50 & -0.05 & 0.05 & 1.04 & -0.24 & 1.73 & -0.37 & 1.90 \\ 
  IE1\_L2 & 0.00 & 0.80 & 250 & -0.05 & -0.03 & 0.57 & -0.07 & 0.41 & -0.12 & 0.42 \\ 
  IE1\_L2 & 0.00 & 0.80 & 500 & -0.05 & -0.06 & 0.42 & -0.07 & 0.26 & -0.10 & 0.27 \\ 
  IE1\_L2 & 0.80 & 0.00 & 50 & 0.05 & 0.41 & 1.04 & 0.01 & 1.77 & -0.12 & 1.76 \\ 
  IE1\_L2 & 0.80 & 0.00 & 250 & 0.05 & 0.22 & 0.56 & 0.04 & 0.48 & 0.03 & 0.52 \\ 
  IE1\_L2 & 0.80 & 0.00 & 500 & 0.05 & 0.16 & 0.45 & -0.01 & 0.30 & 0.01 & 0.35 \\ 
  \hline
  IE2 & 0.00 & 0.00 & 50 & 0.00 & -0.02 & 0.49 & -0.02 & 0.46 & -0.03 & 0.41 \\ 
  IE2 & 0.00 & 0.00 & 250 & 0.00 & 0.01 & 0.16 & 0.00 & 0.15 & 0.00 & 0.14 \\ 
  IE2 & 0.00 & 0.00 & 500 & 0.00 & -0.00 & 0.10 & -0.00 & 0.10 & -0.00 & 0.10 \\ 
  IE2 & 0.00 & 0.80 & 50 & -0.02 & -0.01 & 0.58 & 0.01 & 0.52 & -0.05 & 0.46 \\ 
  IE2 & 0.00 & 0.80 & 250 & -0.02 & -0.02 & 0.27 & 0.01 & 0.16 & -0.05 & 0.17 \\ 
  IE2 & 0.00 & 0.80 & 500 & -0.02 & -0.03 & 0.20 & 0.01 & 0.12 & -0.04 & 0.12 \\ 
  IE2 & 0.80 & 0.00 & 50 & 0.02 & -0.01 & 0.44 & -0.03 & 0.62 & -0.04 & 0.47 \\ 
  IE2 & 0.80 & 0.00 & 250 & 0.02 & -0.00 & 0.18 & 0.00 & 0.16 & -0.00 & 0.16 \\ 
  IE2 & 0.80 & 0.00 & 500 & 0.02 & -0.00 & 0.12 & -0.00 & 0.11 & -0.01 & 0.11 \\ 
  \hline
  IE2\_L2 & 0.00 & 0.00 & 50 & -0.02 & 0.07 & 0.61 & 0.01 & 0.69 & 0.00 & 0.64 \\ 
  IE2\_L2 & 0.00 & 0.00 & 250 & -0.02 & 0.01 & 0.24 & 0.00 & 0.23 & 0.00 & 0.21 \\ 
  IE2\_L2 & 0.00 & 0.00 & 500 & -0.02 & 0.00 & 0.15 & 0.00 & 0.16 & 0.00 & 0.14 \\ 
  IE2\_L2 & 0.00 & 0.80 & 50 & -0.36 & -0.12 & 0.70 & -0.26 & 0.90 & -0.30 & 0.80 \\ 
  IE2\_L2 & 0.00 & 0.80 & 250 & -0.36 & -0.34 & 0.44 & -0.45 & 0.32 & -0.46 & 0.27 \\ 
  IE2\_L2 & 0.00 & 0.80 & 500 & -0.36 & -0.32 & 0.35 & -0.42 & 0.21 & -0.42 & 0.19 \\ 
  IE2\_L2 & 0.80 & 0.00 & 50 & 0.56 & 0.47 & 0.88 & 0.50 & 0.89 & 0.26 & 0.85 \\ 
  IE2\_L2 & 0.80 & 0.00 & 250 & 0.56 & 0.49 & 0.41 & 0.67 & 0.31 & 0.58 & 0.33 \\ 
  IE2\_L2 & 0.80 & 0.00 & 500 & 0.56 & 0.50 & 0.33 & 0.67 & 0.21 & 0.57 & 0.23 \\ 
  \hline
  Joint IE & 0.00 & 0.00 & 50 & 0.02 & -0.02 & 0.51 & -0.05 & 0.63 & -0.04 & 0.63 \\ 
  Joint IE & 0.00 & 0.00 & 250 & 0.02 & -0.00 & 0.17 & -0.00 & 0.17 & 0.00 & 0.19 \\ 
  Joint IE & 0.00 & 0.00 & 500 & 0.02 & -0.00 & 0.11 & -0.00 & 0.11 & -0.00 & 0.13 \\ 
  Joint IE & 0.00 & 0.80 & 50 & -0.17 & -0.15 & 0.69 & -0.04 & 0.86 & -0.16 & 0.92 \\ 
  Joint IE & 0.00 & 0.80 & 250 & -0.17 & -0.15 & 0.34 & 0.02 & 0.18 & -0.18 & 0.23 \\ 
  Joint IE & 0.00 & 0.80 & 500 & -0.17 & -0.17 & 0.26 & 0.03 & 0.12 & -0.18 & 0.17 \\ 
  Joint IE & 0.80 & 0.00 & 50 & 0.01 & -0.01 & 0.62 & -0.04 & 0.69 & -0.03 & 0.72 \\ 
  Joint IE & 0.80 & 0.00 & 250 & 0.01 & -0.01 & 0.19 & -0.02 & 0.17 & -0.03 & 0.22 \\ 
  Joint IE & 0.80 & 0.00 & 500 & 0.01 & -0.02 & 0.13 & -0.02 & 0.12 & -0.02 & 0.15 \\ 
  \hline
  Joint IE\_L2 & 0.00 & 0.00 & 50 & -0.14 & -0.14 & 1.13 & -0.32 & 2.33 & -0.44 & 2.11 \\ 
  Joint IE\_L2 & 0.00 & 0.00 & 250 & -0.14 & -0.08 & 0.42 & -0.11 & 0.37 & -0.12 & 0.39 \\ 
  Joint IE\_L2 & 0.00 & 0.00 & 500 & -0.14 & -0.10 & 0.29 & -0.11 & 0.24 & -0.11 & 0.25 \\ 
  Joint IE\_L2 & 0.00 & 0.80 & 50 & -0.53 & -0.38 & 1.08 & -0.51 & 1.98 & -0.71 & 1.81 \\ 
  Joint IE\_L2 & 0.00 & 0.80 & 250 & -0.53 & -0.52 & 0.51 & -0.50 & 0.34 & -0.65 & 0.37 \\ 
  Joint IE\_L2 & 0.00 & 0.80 & 500 & -0.53 & -0.51 & 0.41 & -0.47 & 0.24 & -0.61 & 0.26 \\ 
  Joint IE\_L2 & 0.80 & 0.00 & 50 & 0.59 & 0.50 & 0.92 & 0.43 & 1.91 & 0.18 & 1.80 \\ 
  Joint IE\_L2 & 0.80 & 0.00 & 250 & 0.59 & 0.58 & 0.45 & 0.58 & 0.36 & 0.60 & 0.41 \\ 
  Joint IE\_L2 & 0.80 & 0.00 & 500 & 0.59 & 0.57 & 0.33 & 0.56 & 0.23 & 0.58 & 0.29 \\
  \hline 
  Mutual IE & 0.00 & 0.00 & 50 & 0.01 & -0.02 & 0.45 & -0.07 & 0.53 & -0.06 & 0.57 \\ 
  Mutual IE & 0.00 & 0.00 & 250 & 0.01 & -0.01 & 0.11 & -0.01 & 0.10 & -0.00 & 0.14 \\ 
  Mutual IE & 0.00 & 0.00 & 500 & 0.01 & -0.00 & 0.06 & -0.01 & 0.06 & -0.00 & 0.08 \\ 
  Mutual IE & 0.00 & 0.80 & 50 & -0.16 & -0.13 & 0.64 & -0.07 & 0.78 & -0.13 & 0.84 \\ 
  Mutual IE & 0.00 & 0.80 & 250 & -0.16 & -0.12 & 0.31 & 0.01 & 0.13 & -0.13 & 0.17 \\ 
  Mutual IE & 0.00 & 0.80 & 500 & -0.16 & -0.13 & 0.23 & 0.02 & 0.07 & -0.14 & 0.12 \\ 
  Mutual IE & 0.80 & 0.00 & 50 & -0.02 & -0.03 & 0.56 & -0.07 & 0.63 & -0.02 & 0.62 \\ 
  Mutual IE & 0.80 & 0.00 & 250 & -0.02 & -0.04 & 0.25 & -0.05 & 0.11 & -0.05 & 0.18 \\ 
  Mutual IE & 0.80 & 0.00 & 500 & -0.02 & -0.03 & 0.20 & -0.04 & 0.07 & -0.03 & 0.12 \\ 
  \hline
  Mutual IE\_L2 & 0.00 & 0.00 & 50 & 0.01 & -0.23 & 0.99 & -0.03 & 1.17 & 0.01 & 0.99 \\ 
  Mutual IE\_L2 & 0.00 & 0.00 & 250 & 0.01 & -0.05 & 0.29 & -0.04 & 0.20 & -0.01 & 0.21 \\ 
  Mutual IE\_L2 & 0.00 & 0.00 & 500 & 0.01 & -0.03 & 0.19 & -0.03 & 0.13 & -0.01 & 0.13 \\ 
  Mutual IE\_L2 & 0.00 & 0.80 & 50 & -0.12 & -0.31 & 1.05 & -0.01 & 1.26 & -0.04 & 1.12 \\ 
  Mutual IE\_L2 & 0.00 & 0.80 & 250 & -0.12 & -0.15 & 0.53 & 0.02 & 0.24 & -0.08 & 0.26 \\ 
  Mutual IE\_L2 & 0.00 & 0.80 & 500 & -0.12 & -0.14 & 0.41 & 0.02 & 0.14 & -0.09 & 0.17 \\ 
  Mutual IE\_L2 & 0.80 & 0.00 & 50 & -0.01 & -0.38 & 1.29 & -0.08 & 1.01 & 0.04 & 0.82 \\ 
  Mutual IE\_L2 & 0.80 & 0.00 & 250 & -0.01 & -0.14 & 0.46 & -0.13 & 0.24 & -0.02 & 0.26 \\ 
  Mutual IE\_L2 & 0.80 & 0.00 & 500 & -0.01 & -0.09 & 0.39 & -0.10 & 0.15 & -0.00 & 0.17 \\ 
  \hline
  DE & 0.00 & 0.00 & 50 & -0.00 & 0.01 & 2.10 & -0.07 & 1.80 & -0.08 & 1.67 \\ 
  DE & 0.00 & 0.00 & 250 & -0.00 & -0.01 & 0.45 & -0.01 & 0.44 & -0.01 & 0.42 \\ 
  DE & 0.00 & 0.00 & 500 & -0.00 & -0.00 & 0.31 & -0.01 & 0.30 & -0.01 & 0.30 \\ 
  DE & 0.00 & 0.80 & 50 & -0.01 & -0.24 & 2.81 & -0.22 & 1.83 & -0.09 & 1.62 \\ 
  DE & 0.00 & 0.80 & 250 & -0.01 & -0.04 & 0.55 & -0.21 & 0.44 & 0.01 & 0.43 \\ 
  DE & 0.00 & 0.80 & 500 & -0.01 & -0.03 & 0.39 & -0.24 & 0.31 & -0.02 & 0.31 \\ 
  DE & 0.80 & 0.00 & 50 & -0.00 & 0.04 & 2.52 & 0.01 & 1.82 & -0.00 & 1.74 \\ 
  DE & 0.80 & 0.00 & 250 & -0.00 & -0.02 & 0.50 & -0.03 & 0.46 & -0.03 & 0.45 \\ 
  DE & 0.80 & 0.00 & 500 & -0.00 & -0.01 & 0.33 & -0.02 & 0.31 & -0.02 & 0.30 \\ 
   \hline\hline
\end{longtable}
\endgroup

\clearpage

\section{Applied example}\label{sect:appliedex:regcoef}

\begin{table}[ht]
\centering
\caption{Regression coefficient estimates from the fitted logistic propensity score model.}
\begin{tabular}{rrrrr}
  \hline
 & Estimate & Std. Error & z value & Pr($>$$|$z$|$) \\ 
  \hline
(Intercept) & -2.37 & 0.71 & -3.36 & 0.00 \\ 
  age & 0.08 & 0.02 & 4.06 & 0.00 \\ 
  gender & -0.03 & 0.20 & -0.15 & 0.88 \\ 
  years\_working & -0.46 & 0.30 & -1.56 & 0.12 \\ 
  marital & 0.02 & 0.26 & 0.07 & 0.95 \\ 
  edu & 0.43 & 0.22 & 2.00 & 0.05 \\ 
  rank & -0.42 & 0.20 & -2.08 & 0.04 \\ 
  neg\_cop & -0.04 & 0.09 & -0.48 & 0.63 \\ 
   \hline
\end{tabular}
\end{table}

\begingroup
\renewcommand\arraystretch{0.7}
\begin{longtable}{rrrrr}
\caption{Regression coefficient estimates from the fitted linear mediator models.}\\
  \hline
 & Estimate & Std. Error & t value & Pr($>$$|$t$|$) \\ 
  \hline\hline\endfirsthead
  \multicolumn{5}{c}{$\E(M_1 | A=0,L)$} \\
  \hline  
(Intercept) & 11.51 & 2.12 & 5.44 & 0.00 \\ 
  age & -0.06 & 0.06 & -0.94 & 0.35 \\ 
  gender & 0.98 & 0.63 & 1.55 & 0.12 \\ 
  years\_working & 0.40 & 0.85 & 0.47 & 0.64 \\ 
  marital & -0.83 & 0.76 & -1.08 & 0.28 \\ 
  edu & -0.65 & 0.64 & -1.02 & 0.31 \\ 
  rank & 1.22 & 0.64 & 1.90 & 0.06 \\ 
  neg\_cop & 1.61 & 0.29 & 5.57 & 0.00 \\  
   \hline
  \multicolumn{5}{c}{$\E(M_1 | A=1,L)$} \\   
  \hline
(Intercept) & 7.96 & 1.65 & 4.82 & 0.00 \\ 
  age & 0.00 & 0.04 & 0.09 & 0.93 \\ 
  gender & 0.40 & 0.48 & 0.84 & 0.40 \\ 
  years\_working & 0.07 & 0.74 & 0.09 & 0.92 \\ 
  marital & -1.01 & 0.63 & -1.60 & 0.11 \\ 
  edu & 0.60 & 0.53 & 1.14 & 0.26 \\ 
  rank & 0.08 & 0.45 & 0.18 & 0.86 \\ 
  neg\_cop & 0.94 & 0.21 & 4.59 & 0.00 \\ 
   \hline
  \multicolumn{5}{c}{$\E(M_2 | A=0, M_1,L)$} \\   
   \hline
(Intercept) & 5.85 & 1.71 & 3.43 & 0.00 \\ 
  M1 & 0.95 & 0.05 & 17.69 & 0.00 \\ 
  neg\_cop & 0.87 & 0.51 & 1.70 & 0.09 \\ 
  age & -0.03 & 0.05 & -0.73 & 0.47 \\ 
  gender & -0.87 & 0.48 & -1.80 & 0.07 \\ 
  years\_working & -0.14 & 0.65 & -0.22 & 0.83 \\ 
  marital & -0.88 & 0.58 & -1.52 & 0.13 \\ 
  edu & -0.76 & 0.49 & -1.56 & 0.12 \\ 
  rank & 0.16 & 0.49 & 0.32 & 0.75 \\ 
  M1:neg\_cop & -0.03 & 0.05 & -0.61 & 0.54 \\  
   \hline
  \multicolumn{5}{c}{$\E(M_2 | A=1, M_1,L)$} \\   
  \hline
(Intercept) & 3.20 & 1.21 & 2.65 & 0.01 \\ 
  M1 & 0.89 & 0.04 & 20.79 & 0.00 \\ 
  neg\_cop & -1.42 & 0.37 & -3.81 & 0.00 \\ 
  age & 0.01 & 0.03 & 0.30 & 0.76 \\ 
  gender & 0.07 & 0.34 & 0.20 & 0.85 \\ 
  years\_working & -0.66 & 0.52 & -1.27 & 0.21 \\ 
  marital & 0.18 & 0.44 & 0.41 & 0.68 \\ 
  edu & -0.58 & 0.37 & -1.55 & 0.12 \\ 
  rank & 0.16 & 0.32 & 0.51 & 0.61 \\ 
  M1:neg\_cop & 0.23 & 0.04 & 5.48 & 0.00 \\ 
   \hline
  \multicolumn{5}{c}{$\E(M_3 | A=0, M_1, M_2, L)$} \\   
  \hline
(Intercept) & -1.81 & 1.73 & -1.04 & 0.30 \\ 
  M1 & 0.27 & 0.12 & 2.23 & 0.03 \\ 
  M2 & 0.79 & 0.10 & 7.75 & 0.00 \\ 
  neg\_cop & 1.15 & 0.98 & 1.16 & 0.25 \\ 
  age & 0.03 & 0.04 & 0.91 & 0.36 \\ 
  gender & 0.45 & 0.38 & 1.18 & 0.24 \\ 
  years\_working & -0.14 & 0.51 & -0.27 & 0.78 \\ 
  marital & -0.54 & 0.46 & -1.17 & 0.24 \\ 
  edu & 0.39 & 0.38 & 1.01 & 0.31 \\ 
  rank & -0.70 & 0.39 & -1.82 & 0.07 \\ 
  M1:M2 & -0.01 & 0.01 & -1.63 & 0.10 \\ 
  M1:neg\_cop & -0.02 & 0.11 & -0.22 & 0.83 \\ 
  M2:neg\_cop & -0.13 & 0.09 & -1.43 & 0.16 \\ 
  M1:M2:neg\_cop & 0.01 & 0.01 & 1.21 & 0.23 \\  
   \hline
  \multicolumn{5}{c}{$\E(M_3 | A=1, M_1, M_2,L)$} \\  
  \hline
(Intercept) & 2.41 & 1.63 & 1.48 & 0.14 \\ 
  M1 & 0.28 & 0.13 & 2.13 & 0.03 \\ 
  M2 & 0.55 & 0.13 & 4.36 & 0.00 \\ 
  neg\_cop & 0.78 & 1.07 & 0.73 & 0.47 \\ 
  age & -0.04 & 0.03 & -1.50 & 0.13 \\ 
  gender & -0.12 & 0.33 & -0.36 & 0.72 \\ 
  years\_working & 0.58 & 0.51 & 1.15 & 0.25 \\ 
  marital & -0.03 & 0.43 & -0.06 & 0.95 \\ 
  edu & -0.27 & 0.37 & -0.75 & 0.45 \\ 
  rank & 0.29 & 0.31 & 0.94 & 0.35 \\ 
  M1:M2 & -0.01 & 0.01 & -0.77 & 0.44 \\ 
  M1:neg\_cop & -0.07 & 0.13 & -0.53 & 0.60 \\ 
  M2:neg\_cop & -0.08 & 0.13 & -0.63 & 0.53 \\ 
  M1:M2:neg\_cop & 0.01 & 0.01 & 0.89 & 0.38 \\ 
   \hline
\end{longtable}
\endgroup

\begin{table}[ht]
\centering
\caption{Regression coefficient estimates from the fitted logistic outcome model.}
\begin{tabular}{rrrrr}
  \hline
 & Estimate & Std. Error & z value & Pr($>$$|$z$|$) \\ 
  \hline\hline
  \multicolumn{5}{c}{$\logit\{\Pr(Y=1 | A=0, M_1, M_2, M_3,L)\}$} \\    
  \hline  
(Intercept) & -3.93 & 2.06 & -1.90 & 0.06 \\ 
  M1 & 0.27 & 0.24 & 1.15 & 0.25 \\ 
  M2 & 0.35 & 0.24 & 1.44 & 0.15 \\ 
  M3 & 0.09 & 0.25 & 0.34 & 0.73 \\ 
  age & -0.01 & 0.04 & -0.33 & 0.74 \\ 
  gender & -0.41 & 0.41 & -1.00 & 0.32 \\ 
  years\_working & -0.17 & 0.57 & -0.30 & 0.76 \\ 
  marital & 0.32 & 0.47 & 0.68 & 0.50 \\ 
  edu & -0.42 & 0.41 & -1.04 & 0.30 \\ 
  rank & 0.50 & 0.42 & 1.19 & 0.23 \\ 
  neg\_cop & -0.63 & 1.26 & -0.50 & 0.62 \\ 
  M1:M2 & -0.03 & 0.02 & -1.60 & 0.11 \\ 
  M1:M3 & 0.01 & 0.03 & 0.44 & 0.66 \\ 
  M2:M3 & 0.00 & 0.02 & 0.12 & 0.90 \\ 
  M1:neg\_cop & 0.30 & 0.24 & 1.24 & 0.21 \\ 
  M2:neg\_cop & -0.24 & 0.24 & -0.99 & 0.32 \\ 
  M3:neg\_cop & 0.06 & 0.26 & 0.22 & 0.82 \\ 
  M1:M2:neg\_cop & -0.00 & 0.02 & -0.11 & 0.92 \\ 
  M1:M3:neg\_cop & -0.03 & 0.03 & -1.16 & 0.25 \\ 
  M2:M3:neg\_cop & 0.03 & 0.02 & 1.25 & 0.21 \\  
   \hline
  \multicolumn{5}{c}{$\logit\{\Pr(Y=1 | A=1, M_1, M_2, M_3,L)\}$} \\    
  \hline
(Intercept) & -6.11 & 1.88 & -3.25 & 0.00 \\ 
  M1 & 0.13 & 0.24 & 0.53 & 0.60 \\ 
  M2 & 0.30 & 0.23 & 1.28 & 0.20 \\ 
  M3 & 0.57 & 0.21 & 2.70 & 0.01 \\ 
  age & -0.03 & 0.03 & -0.82 & 0.41 \\ 
  gender & 0.24 & 0.34 & 0.68 & 0.49 \\ 
  years\_working & 0.33 & 0.52 & 0.64 & 0.52 \\ 
  marital & 0.07 & 0.45 & 0.16 & 0.87 \\ 
  edu & -0.27 & 0.37 & -0.73 & 0.46 \\ 
  rank & -0.04 & 0.32 & -0.12 & 0.90 \\ 
  neg\_cop & 0.48 & 1.42 & 0.34 & 0.74 \\ 
  M1:M2 & -0.01 & 0.02 & -0.64 & 0.52 \\ 
  M1:M3 & -0.00 & 0.02 & -0.04 & 0.97 \\ 
  M2:M3 & -0.02 & 0.02 & -0.75 & 0.45 \\ 
  M1:neg\_cop & 0.18 & 0.31 & 0.59 & 0.55 \\ 
  M2:neg\_cop & -0.19 & 0.28 & -0.69 & 0.49 \\ 
  M3:neg\_cop & -0.08 & 0.23 & -0.33 & 0.74 \\ 
  M1:M2:neg\_cop & 0.01 & 0.02 & 0.52 & 0.60 \\ 
  M1:M3:neg\_cop & -0.03 & 0.03 & -0.89 & 0.37 \\ 
  M2:M3:neg\_cop & 0.02 & 0.03 & 0.69 & 0.49 \\ 
   \hline
\end{tabular}
\end{table}

\end{document}